\documentstyle[12pt,epsf]{article}

\def\thefootnote{\fnsymbol{footnote}}

\begin{document}
\begin{titlepage}
\begin{center}
{\Large\bf Aspects of $W$ Physics at the Linear Collider}
\footnote{Plenary Talk, given at LCW95, Morioka-Appi (Japan), Sep. 8-12 1995.\\
In the proceedings of the Workshop {\em Physics and Experiments with Linear Colliders}, Vol.~1, 
p.~199-227. Edited by A.~Miyamoto, Y.~Fujii, T.~Matsui and S.~Iwata, World Scientific, 1996.}

{\bf F.~Boudjema }
\\
{\it Laboratoire de Physique Th\'eorique} 
EN{\large S}{\Large L}{\large A}PP
\footnote{URA 14-36 du CNRS, associ\'ee \`a l'E.N.S de Lyon et \`a
l'Universit\'e de Savoie.}\\
\smallskip
{\it Chemin de Bellevue, B.P. 110, F-74941 Annecy-le-Vieux, C\'edex, France.}
\end{center}
\vspace*{\fill}
\centerline{{\bf Abstract}}

\baselineskip=14pt
\noindent
%%%%%%%%%%%%%%%%%%%%%%%%%%%%%%%%%%%%%%%%%%%%%%%%%%%%%%%%%%%%%%%%%%%%%
\small 
It is argued that the next linear colliders can serve as 
$W$ factories that may be exploited for precision tests on 
the properties of the massive gauge bosons. The connection between 
the probing of the symmetry breaking sector and precise measurement
of the self-couplings of the weak vector bosons is stressed. The 
discussion relies much on the impact of the present low-energy data, 
especially 
LEP1. These have restricted some paths that lead to the exploration of 
the scalar sector through the investigation of the so-called 
anomalous couplings of the $W$'s and suggest a hierarchy 
in the classification of these parameters. 
The limits we expect to set on these 
couplings at the different modes of the linear colliders 
 are reviewed and compared with those one obtains at the LHC. The conclusion 
is that the first phase of a linear collider running at 
500 GeV and the LHC are complementary. Some important issues concerning 
radiative corrections and backgrounds that need further studies 
in order that one conducts high precision analyses at high energies 
are discussed. 

\normalsize \protect \normalsize

\vspace*{\fill}
\vspace*{0.1cm}
\rightline{ENSLAPP-A-575/96} 
%\rightline{hep-ph/yymmxxx}
\rightline{January 1996}
\vspace*{\fill}
%\noindent 
%{\footnotesize $\;^{*}$ URA 14-36 du CNRS, associ\'ee \`a l'E.N.S de
%Lyon et \`a l'Universit\'e de Savoie.}
\end{titlepage}
\baselineskip=18pt
\font\tenbf=cmbx10
\font\tenrm=cmr10
\font\tenit=cmti10
\font\elevenbf=cmbx10 scaled\magstep 1
\font\elevenrm=cmr10 scaled\magstep 1
\font\elevenit=cmti10 scaled\magstep 1
\font\ninebf=cmbx9
\font\ninerm=cmr9
\font\nineit=cmti9
\font\eightbf=cmbx8
\font\eightrm=cmr8
\font\eightit=cmti8
\font\sevenrm=cmr7
\oddsidemargin  0in
\evensidemargin 0in
\def\thefootnote{\fnsymbol{footnote}}

\textwidth 6.0in
\textheight 8.5in
\topmargin -0.25truein
\oddsidemargin 0.30truein
\evensidemargin 0.30truein
\raggedbottom
\parindent=3pc
\baselineskip=10pt
%\begin{document}
%\hfill July 1993
%\hfill ENSLAPP-A-431/93\\
\begin{center}{
{\bf Aspects of $W$ Physics at the Linear Collider}
%\footnote{\ninerm Invited Talk given at the ``Beyond the Standard Model IV,
%12-16 Dec. 1994.\\}
\vglue 5pt
{\rm Fawzi Boudjema\\}
\baselineskip=13pt
{\tenit Laboratoire de Physique Th\'eorique} 
EN{\large S}{\Large L}{\large A}PP
\footnote{\ninerm URA 14-36 du CNRS, associ\'ee \`a l'E.N.S de Lyon 
et \`a l'Universit\'e de Savoie}\\ 
{\tenit Chemin de Bellevue, B.P. 110, F-74941 Annecy-le-Vieux, Cedex, France.}\\
{\tenrm E-mail:BOUDJEMA@LAPPHP8.IN2P3.FR}
\vglue 0.3cm
{\tenrm ABSTRACT}} \\
\end{center} 
\noindent 
\small 
It is argued that the next linear colliders can serve as 
$W$ factories that may be exploited for precision tests on 
the properties of the massive gauge bosons. The connection between 
the probing of the symmetry breaking sector and precise measurement
of the self-couplings of the weak vector bosons is stressed. The 
discussion relies much on the impact of the present low-energy data, 
especially 
LEP1. These have restricted some paths that lead to the exploration of 
the scalar sector through the investigation of the so-called 
anomalous couplings of the $W$'s and suggest a hierarchy 
in the classification of these parameters. 
The limits we expect to set on these 
couplings at the different modes of the linear colliders 
 are reviewed and compared with those one obtains at the LHC. The conclusion 
is that the first phase of a linear collider running at 
500 GeV and the LHC are complementary. Some important issues concerning 
radiative corrections and backgrounds that need further studies 
in order that one conducts high precision analyses at high energies 
are discussed. 

\normalsize \protect \normalsize

\vglue 0.3cm
{\rightskip=3pc
 \leftskip=3pc
 \tenrm\baselineskip=12pt
 \noindent

\vglue 0.6cm}
\baselineskip=14pt

\newcommand{\beq}{\begin{equation}}
\newcommand{\eeq}{\end{equation}}

\newcommand{\beqn}{\begin{eqnarray}}
\newcommand{\eeqn}{\end{eqnarray}}
 
\newcommand{\ra}{\rightarrow}
 
\newcommand{\su}{$ SU(2) \times U(1)\,$}
 
\newcommand{\gag}{$\gamma \gamma$ }
\newcommand{\gam}{\gamma \gamma }

\newcommand{\np}{Nucl.\,Phys.\,}
\newcommand{\pl}{Phys.\,Lett.\,}
\newcommand{\pr}{Phys.\,Rev.\,}
\newcommand{\prl}{Phys.\,Rev.\,Lett.\,}
\newcommand{\prep}{Phys.\,Rep.\,}
\newcommand{\zp}{Z.\,Phys.\,}
\newcommand{\sovjnp}{{\em Sov.\ J.\ Nucl.\ Phys.\ }}
\newcommand{\nuclinst}{{\em Nucl.\ Instrum.\ Meth.\ }}
\newcommand{\annp}{{\em Ann.\ Phys.\ }}
\newcommand{\intjmp}{{\em Int.\ J.\ of Mod.\ Phys.\ }}
 
\newcommand{\eps}{\epsilon}
\newcommand{\mw}{M_{W}}
\newcommand{\mww}{M_{W}^{2}}
\newcommand{\mwmw}{M_{W}^{2}}
\newcommand{\mhmh}{M_{H}^2}
\newcommand{\mz}{M_{Z}}
\newcommand{\mzz}{M_{Z}^{2}}

\newcommand{\lra}{\leftrightarrow}
\newcommand{\tr}{{\rm Tr}}

\newcommand{\dkg}{\Delta \kappa_{\gamma}}
\newcommand{\dkz}{\Delta \kappa_{Z}}
\newcommand{\dz}{\delta_{Z}}
\newcommand{\dgz}{\Delta g^{1}_{Z}}
\newcommand{\dgzt}{$\Delta g^{1}_{Z}\;$}
\newcommand{\la}{\lambda}
\newcommand{\lag}{\lambda_{\gamma}}
\newcommand{\laz}{\lambda_{Z}}
\newcommand{\lnl}{L_{9L}}
\newcommand{\lnr}{L_{9R}}
\newcommand{\lt}{L_{10}}
\newcommand{\lu}{L_{1}}
\newcommand{\ld}{L_{2}}

\newcommand{\ememt}{$e^{-} e^{-}\;$}
\newcommand{\epemt}{$e^{+} e^{-}\;$}
\newcommand{\epem}{e^{+} e^{-}\;}
\newcommand{\epemww}{e^{+} e^{-} \ra W^+ W^- \;}
\newcommand{\eewwt}{$e^{+} e^{-} \ra W^+ W^- \;$}
\newcommand{\epemwwt}{$e^{+} e^{-} \ra W^+ W^- \;$}
\newcommand{\ppwg}{p p \ra W \gamma}
\newcommand{\ppwz}{pp \ra W Z}
\newcommand{\ppwgt}{$p p \ra W \gamma \;$}
\newcommand{\ppwzt}{$pp \ra W Z \;$}
\newcommand{\gamgamt}{$\gamma \gamma \;$}
\newcommand{\gamgam}{\gamma \gamma \;}
\newcommand{\egamt}{$e \gamma \;$}
\newcommand{\egam}{e \gamma \;}
\newcommand{\gamgamwwt}{$\gamma \gamma \ra W^+ W^- \;$}
\newcommand{\gamgamwwht}{$\gamma \gamma \ra W^+ W^- H \;$}
\newcommand{\gamgamwwh}{\gamma \gamma \ra W^+ W^- H \;}

\newcommand{\ptu}{p_{1\bot}}
\newcommand{\vecptu}{\vec{p}_{1\bot}}
\newcommand{\ptd}{p_{2\bot}}
\newcommand{\vecptd}{\vec{p}_{2\bot}}
\newcommand{\ie}{{\em i.e.}}
\newcommand{\cm}{{{\cal M}}}
\newcommand{\cl}{{{\cal L}}}
\newcommand{\cd}{{{\cal D}}}
\newcommand{\cv}{{{\cal V}}}
\def\slashc{c\kern -.400em {/}}
\def\slashL{L\kern -.450em {/}}
\def\slashcl{\cl\kern -.600em {/}}
\def\W{{\mbox{\boldmath $W$}}}
\def\B{{\mbox{\boldmath $B$}}}
\def\noi{\noindent}
\def\nn{\noindent}
\def\sm{${\cal{S}} {\cal{M}}\;$}
\def\nph{${\cal{N}} {\cal{P}}\;$}
\def\sb{$ {\cal{S}} {\cal{B}}\;$}
\def\ssb{${\cal{S}} {\cal{S}} {\cal{B}}\;$}
\def\cviol{${\cal{C}}\;$}
\def\pviol{${\cal{P}}\;$}
\def\cpviol{${\cal{C}} {\cal{P}}\;$}

\newcommand{\lgg}{\lambda_1\lambda_2}
\newcommand{\lww}{\lambda_3\lambda_4}
\newcommand{\ppin}{ P^+_{12}}
\newcommand{\pmin}{ P^-_{12}}
\newcommand{\ppout}{ P^+_{34}}
\newcommand{\pmout}{ P^-_{34}}
\newcommand{\sinsq}{\sin^2\theta}
\newcommand{\cossq}{\cos^2\theta}
\newcommand{\yt}{y_\theta}
\newcommand{\hppll}{++;00}
\newcommand{\hpmll}{+-;00}
\newcommand{\hpplt}{++;\lambda_30}
\newcommand{\hpmlt}{+-;\lambda_30}
\newcommand{\hpptt}{++;\lambda_3\lambda_4}
\newcommand{\hpmtt}{+-;\lambda_3\lambda_4} 
\newcommand{\dk}{\Delta\kappa}
\newcommand{\klam}{\Delta\kappa \lambda_\gamma }
\newcommand{\kac}{\Delta\kappa^2 }
\newcommand{\lac}{\lambda_\gamma^2 }
\def\gamgamtzz{$\gamma \gamma \ra ZZ \;$}
\def\gamgamtww{$\gamma \gamma \ra W^+ W^-\;$}
\def\gamgamtwwe{\gamma \gamma \ra W^+ W^-}
\def\ggwwt{$\gamma \gamma \ra W^+ W^-\;$}

\section{Introduction:}
\begin{figure*}[htbp]
\caption{\label{wfactory}{\em Typical sizes of cross sections for weak boson production at the linear colliders.}}
\begin{center}
\mbox{}
\hspace*{-25mm}
\mbox{
\mbox{\epsfxsize=90mm\epsfysize=210mm\epsffile[20 79 519 755]{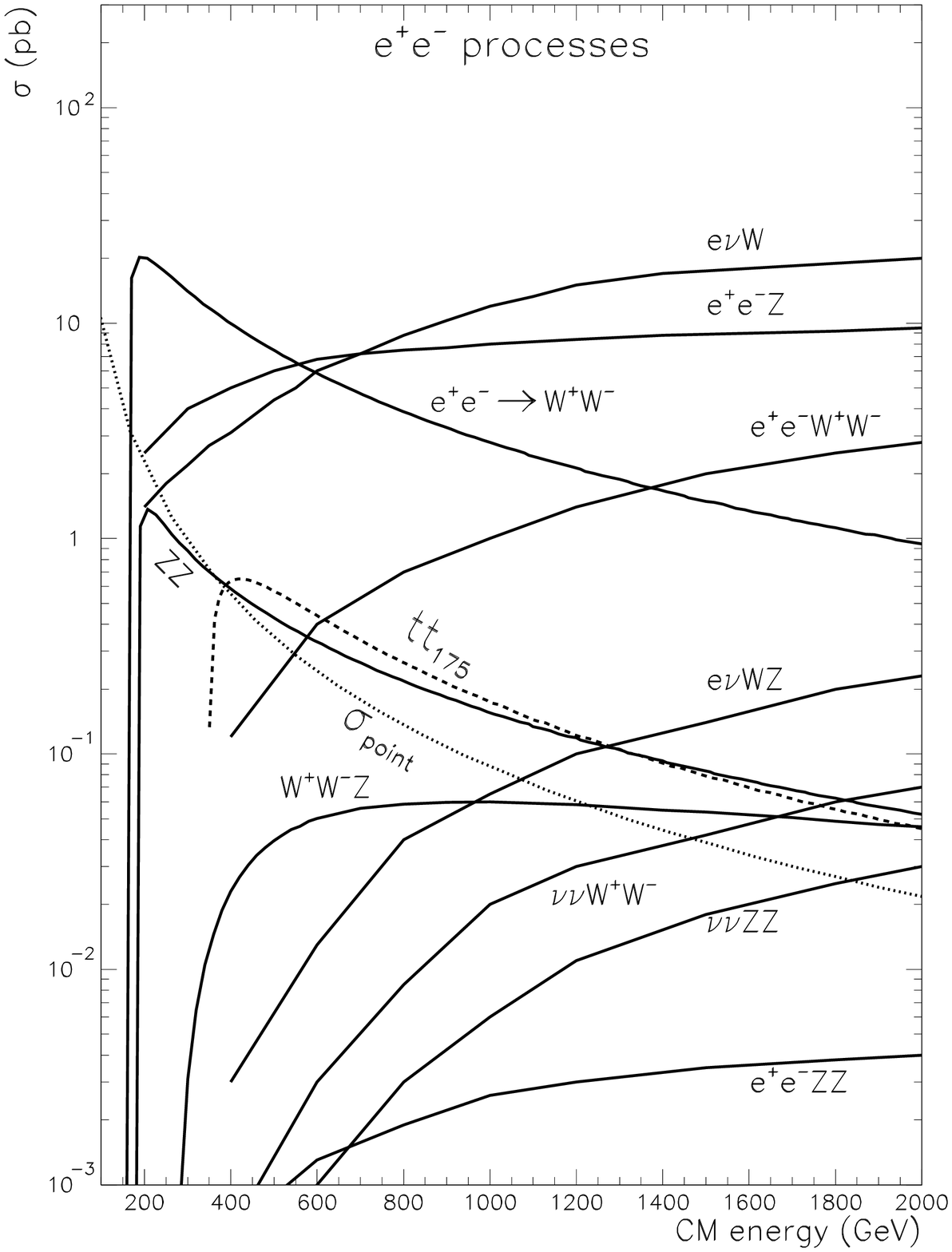}}
\hfill
\mbox{\epsfxsize=90mm\epsfysize=210mm\epsffile[20 79 519 755]{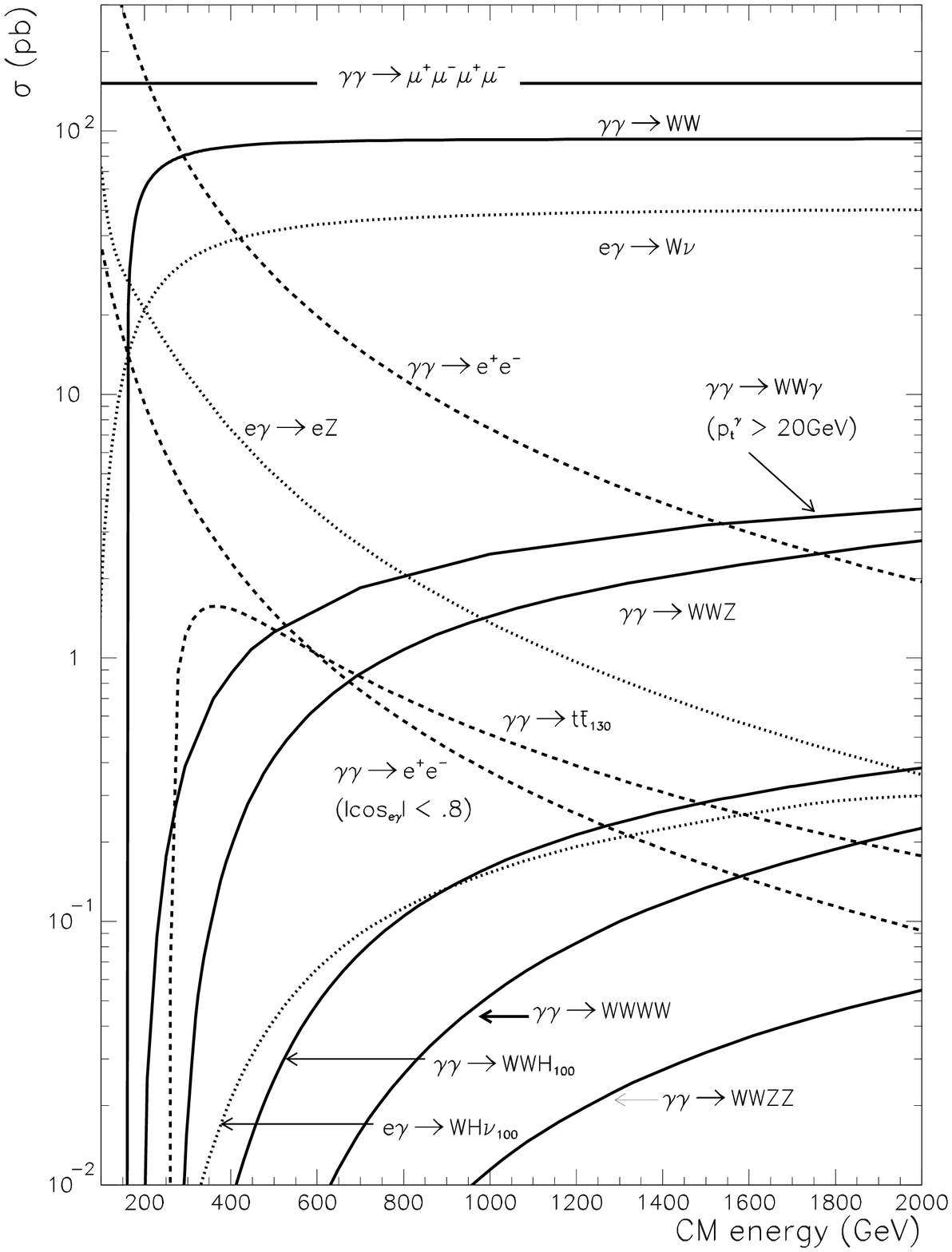}}}
\hspace*{-10mm}
\end{center}
\end{figure*}
\begin{figure*}[htb]
\caption{\label{wlvswt}{\em The proportion of $W_L$ out of 
$W_T$ in \epemwwt and \ggwwt reactions as a function of the centre-of-mass 
energy. How the ratio is slightly improved after cutting the forward events 
is also shown.}}
\begin{center}
\mbox{\epsfxsize=150mm\epsfysize=91mm\epsffile[2 7 558 297]{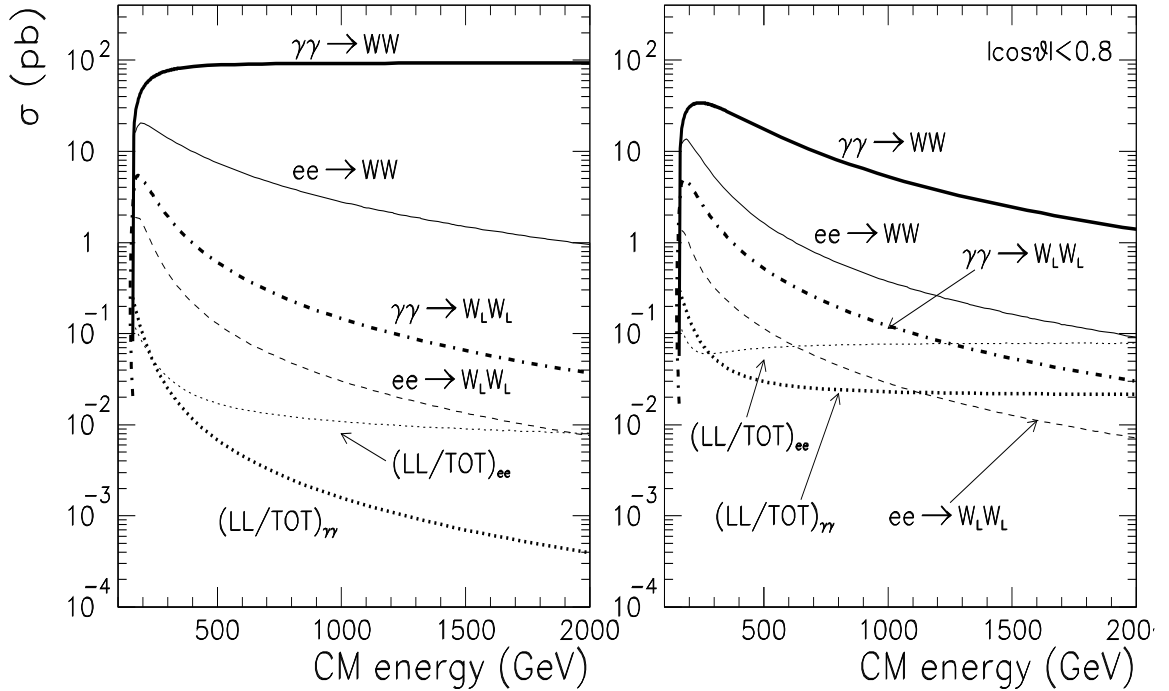}}
\end{center}
\end{figure*}

%\beqn \label{kinetic}
%\cl_G=- \frac{1}{2} \left[ 
%Tr(\W_{\mu \nu} \W^{\mu \nu}) + Tr(\B_{\mu \nu} \B^{\mu \nu}) \right]
%\eeqn
%
The next linear colliders can be regarded as $W$ factories, even and 
especially if run in their \egamt and \gamgamt modes, as evidenced 
from Fig.~\ref{wfactory}. Cross sections for production of 
$Z$ and $W$'s are much larger than the production of fermions and 
scalars, {\it a fortiori\/} larger than the rates of production of 
eventual new particles. Would this be a nuisance or a blessing? 
The answer is intimately related to 
the underlying manifestation of symmetry breaking, 
which is after all the 
{\em raison d'\^etre} of the upcoming high energy machines. 
In a nut-shell one can argue that, if it is supersymmetry that 
solves the naturality problem, then in order to critically 
probe the supersymmetric spectrum, $W$'s will be annoying and in some cases 
potentially dangerous backgrounds. On the other hand if one has to 
live with a heavy Higgs or none, than $W$'s and $Z$'s could play an important 
role through the dynamics of their longitudinal components which is essentially 
the dynamics of the Goldstone Bosons. 
An immediate relevant question is whether these more interesting components are 
produced as numerously as the transverse modes. 
 At this point it is fair to 
stress that, as with all good things, the amount of longitudinals 
out of the very large $W$ production rates is rather small. 
A typical example is given by \epemwwt and \gamgamwwt, see 
Fig.~\ref{wlvswt}. 

These very characteristics of $W$ physics and their impact on the 
search of new physics are set by the fact that 
the massive electroweak bosons constitute a unique system that embodies 
and combines two very fundamental principles: gauge principle and 
symmetry breaking. \\
\noindent 
$\bullet$ The gauge principle accounts for the universality of 
the coupling between all the known elementary fermions and the
weak quanta. In its non-Abelian manifestation this also describes
self-interacting $W$'s with a strength set by the universal coupling. 
These self-interactions, with 
tri-linear and quadri-linear couplings, follow immediately 
from the kinetic terms of $W$'s\footnote{The conventions and definitions of the fields and
matrices that I am using here are the same as those in~\cite{Hawai}.}
\beqn \label{kinetic}
\cl_G=- \frac{1}{2} \left[ 
Tr(\W_{\mu \nu} \W^{\mu \nu}) + Tr(\B_{\mu \nu} \B^{\mu \nu}) \right]
\eeqn
This purely radiation part is present in any unbroken gauge theory
like QCD, say. It describes the propagation and interaction of {\em
transverse} states. 

\noindent 
$\bullet$ 
$\bullet$ An easy and naive way of seeing that a longitudinal 
vector boson does not efficaciously contribute to the purely gauge part
is to observe that the leading component of the polarisation vector 
of a spin-1, the Z say, $\epsilon_\mu$, is $\epsilon_\mu \propto k_\mu/M_Z$ 
($k$ is the 4-momentum of the $Z$). This vector does not contribute to 
the antisymmetric tensor $Z_{\mu \nu}=\partial_\mu Z_\nu - \partial_\nu Z_\mu$. 
The source of this third degree of freedom is the mass term. 
\beqn \label{wmass}
\cl_M=M_W^2 W^+_\mu W^{-\mu} + \frac{1}{2}M_Z^2 Z_\mu Z^\mu
\eeqn
\noindent The mass term would seem to break this crucial local gauge
symmetry. The important point, as you know, is that the symmetry is
not broken but rather hidden. Upon introducing auxiliary fields with
the appropriate gauge transformations, we can rewrite the mass term in
a manifestly local gauge invariant way (through the use of covariant
derivatives). 
In the minimal standard model this is done through a doublet of
scalars, $\Phi$, of which one is the {\em physical\/} Higgs. 
The simplest choice of the doublet implements an extra global {\em
custodial} SU(2) symmetry that gives the well established
$\rho=\frac{M_W^2}{M_Z^2 c_W^2} \simeq 1$. 
\beqn \label{higgspart}
\cl_{H,M}&=&(\cd_\mu \Phi)^\dagger(\cd_\mu \Phi)\;-\;
\lambda \left[ \Phi^\dagger \Phi - \frac{\mu^2}{2 \lambda} 
\right]^2
\eeqn

\noi In the case where the Higgs does not exist or is too heavy, one
can modify this prescription such that only the Goldstone Bosons
$\omega_{1,2,3}$, grouped in the matrix $\Sigma$, are eaten (see for
instance~\cite{Appelquist}): 
\beqn \label{masscov}
\cl_M=\frac{v^2}{4} \tr(\cd^\mu \Sigma^\dagger \cd_\mu \Sigma) \;\;\; ; \;\; 
\Sigma=\exp(\frac{i \omega_{\alpha} \tau^{\alpha}}{v}) \;\;\; (v=246\ GeV)
\eeqn
In this so-called non-linear realisation of \sb 
the mass term~(\ref{wmass}) is formally recovered by going to the physical ``frame" (gauge) 
where all Goldstones disappear, {\it i.e.}, $\Sigma \ra${\bf 1}. 

Viewed this way, the longitudinals are by far the most interesting 
aspect of 
$W$ and $Z$ physics since they are a direct realisation of the Goldstone Bosons
and thus are most likely to shed light on the mechanism of mass. 
Therefore, if one can perform precision measurements on some characteristic properties
of the $W$ and $Z$ interactions one may be able to indirectly reveal the first signs 
of a new interaction that controls symmetry breaking. This could already be done 
at moderate energies 
with processes like \eewwt (or $pp \ra WZ$ at the hadron colliders). 
Ultimately, in the 
event that the Higgs will have been elusive at both the LHC and the 
first stage of 
the NLC one would pursue the investigation of the physics of a strongly interacting 
$W$ at linear colliders in the TeV range through $WW$ scattering 
processes. This second aspect of $W$ dynamics is covered by 
Tim Barklow~\cite{Barklow2}. 

\begin{figure*}[hbt]
\caption{\label{unit.fig}{\em Contribution of the neutrino $t$-channel 
diagrams and the $WW\gamma/Z$ vertices in \epemwwt. }}
%\cite{Barklow}
\begin{center}
\mbox{\mbox{}\hskip-15mm
\mbox{\epsfxsize=150mm\epsfysize=90mm\epsffile[0 20 525 652]{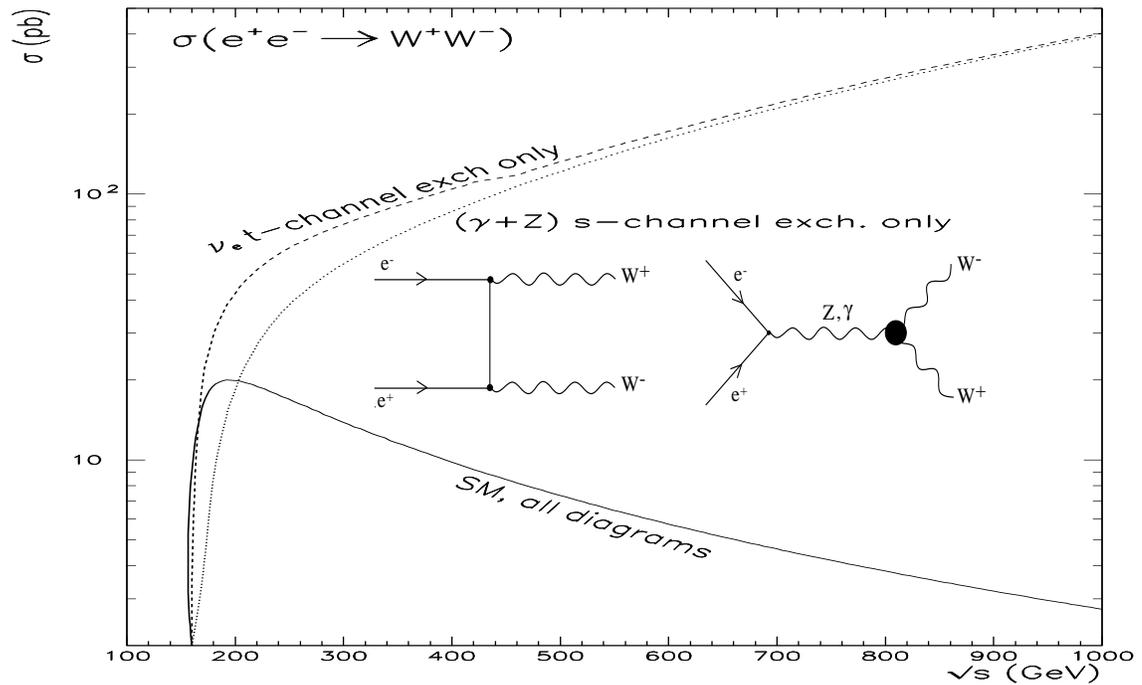}}
\raisebox{3.7cm}{
\mbox{
\hspace*{-10.5cm}\epsfxsize=8cm\epsfysize=2cm\epsffile{figeewwsm.eps}}}}
\end{center}
\end{figure*}

Although the motivation for studying $W$ physics rests essentially on finding 
the underlying 
origin for the mass term, Eq.~\ref{wmass}, of all the pieces 
that build up 
the purely bosonic sector of the \sm Lagrangian, 
the mass terms piece, $M_{W}$ and $M_Z$, (Eq.~\ref{wmass}) 
 is evidently completely established, with its 
parameters very precisely measured. Of course one must interpret this evidence 
as being the tip of whatever iceberg is hidden in the darkness of the 
symmetry 
breaking underworld. 
 On the 
other hand we have not had {\em direct} and precise enough empirical 
evidence for 
all the parts that define the radiation term, namely the piece 
which 
constitutes the hallmark 
of the non-Abelian 
structure: vector bosons self-couplings. 
Without the $WWZ/WW\gamma$ (and the quartic) 
terms the cross sections that are displayed in Fig.~\ref{wfactory}
will get disastrously 
enormous. The text-book example, soon to be 
the {\em bread and butter} of LEP2, is 
\epemwwt. Keeping only the $t$-channel diagram, that involves the well confirmed 
$We\nu_e$ vertex leads to a cross section that gets out of hand, 
and breaks 
unitarity (see Fig.~\ref{unit.fig}). Restoring full gauge invariance 
by the introduction of the \sm vertices for $WW\gamma$ and $WWZ$, the cross
section for $W$ pair production decreases with energy. 
%[0 20 525 652]
%[0 0 1031 224]

It is extremely important to stress that this cancellation is not 
strictly speaking a test of non-Abelian gauge invariance. 
It occurs because one has restricted the $WWZ/WW\gamma$, to their 
\underline{minimal} form which, true, is set by gauge invariance. 
The notion of minimality will be developped further, but one should 
keep in mind that one could have had other forms of the $WWZ/WW\gamma$ 
than those derived from Eq.~\ref{kinetic}, that are perfectly gauge 
invariant but do not necessarily lead to a unitary cross section. 
Minimality means also that one is within the realm of the {\em minimal} 
\sm and therefore 
the probing of these couplings, particularly in the absence of any new particles, notably the 
Higgs, is an indirect way of learning about the New Physics before reaching the scale 
where it is fully manifest. What is more appealing about the 
probing of these couplings is that 
they could really tell us something about the mechanism of symmetry breaking and the dynamics 
of the Goldstone Bosons, which can be regarded as some realisation of the longitudinal part 
of the $W$ and $Z$. Investigating non-minimal values of these couplings means considering 
other structures that involve the interactions of the $W$ that are generalisations 
of the mass operators and the kinetic term operators. It must be 
said that terms that are 
constructed on the mould of the kinetic term (Eq.\ref{kinetic}), even if they lead to 
``anomalous" non-standard values of the tri-linear and quadri-linear 
couplings,
are from the standpoint of symmetry breaking less interesting since they do not involve the 
Goldstone Bosons. I will come back to how a hierarchy of non standard 
 self-couplings 
of the $W$ suggests itself which 
should be used as a guide to concentrate on some couplings 
rather than others. But before doing so I will leave this well 
motivated bias and present, for 
completeness, the most general tri-linear couplings; at least those that do not 
break some well confirmed symmetries such as \cpviol. For a talk on this subject at the 
linear collider it is good to keep in mind the message of Fig.~\ref{unit.fig}: 
going to higher 
centre-of-mass energies really pays. Indeed, as evident from 
Fig.~\ref{unit.fig} a deviation of these couplings from the 
values they attain in the \sm will be magnified as one goes to higher 
energies, and hence more precise tests will be possible. Sadly for LEP2, 
around threshold, the compensation does not require so subtle 
fine tuning.

\section{Anomalous Weak Bosons Self-Couplings: 
Parameterisations and Classifications}

\subsection{The standard ``phenomenological" parameterisation of the tri-linear 
coupling}

One knows~\cite{Majorana}
that a particle of spin-J which is not its own anti-particle 
can have, at most, $(6J+1)$ 
electromagnetic form-factors including \cviol, \pviol and \cpviol 
violating terms. The 
same argument tells us~\cite{Majorana} 
that if the ``scalar"-part of a massive spin-1 
particle does not contribute, as is the case for the Z in 
$e^+ e^- \ra W^+ W^-$, 
then there is also the same number of invariant form-factors for 
the spin-1 coupling to a charged spin-J particle. This means that 
there are 7 independent $WWZ$ form factors and 6 independent $WW\gamma$
form-factors beside the electric charge of the $W$. This number of invariants 
is derived by appealing to angular momentum conservation and to (a lesser 
extent) the 
conservation of the {\em Abelian} $U(1)$ current: {\em i.e.}, {\bf two} utterly 
established symmetry principles one would, at no cost, dare to tamper with. 
Although one can not be more general than this, if all these $13$ 
couplings were simultaneously allowed on the same footing, in an experimental 
fitting procedure and most critically to the best probe $e^+e^- \ra W^+W^-$, 
it will be 
a formidable task to disentangle between all the effects, or to extract 
good limits on all. \\
\noi One then asks 
whether other symmetries, though not as inviolable as the two previous ones, 
may be invoked to reduce 
the set of permitted extra parameters. One expects that the more contrived 
a symmetry has thus far been verified, the less likely a parameter which breaks 
this symmetry is to occur, compared to a parameter which respects these 
symmetries. 
For instance, in view 
of the null results on the electric dipole moments of fermions and other 
\cpviol violating observables pointing to almost no \cpviol violation, 
\cpviol violating terms, and especially the electromagnetic ones, are very 
unlikely to have any detectable 
impact on $W$-pair production. 
Therefore, in a first analysis they should not 
be fitted. The same goes for the \cviol violating $WW\gamma$ couplings. 
Additional symmetry principles and then theoretical ``plausibility 
arguments" can be invoked to further reduce the parameter space of the anomalous
couplings. However, before invoking any additional criteria other than 
angular momentum conservation, conservation of the {\em Abelian} current, 
unobservable \cpviol and electromagnetic \cviol violation, we should give the 
{\em \bf most} general {\em phenomenological} parameterisation of the 
$WWV$ vertex. 
%This parameterisation is to be used at {\em tree-level} in 
%processes describing vector boson pair production by light fermions 
%(or any other crossed channels of these). 
The by-now standard {\em phenomenological} parameterisation of the
$WW\gamma$ and $WWZ$ vertex of HPZH~\cite{HPZH} has been written for
the purpose of studying \eewwt. 
The same parameterisation, although as general as it can be for
\eewwt, may not be necessarily correct nor general when applied to
other situations. 
Nonetheless, the HPZH 
parameterisation has become popular enough in discussing anomalies that I will refer 
to it quite often as a common ground when comparing various
approaches and ``data". 
It assumes the vector bosons to be 
either {\em on-shell} or associated to a conserved current. 
With this warning, one may well be tempted (if all what one cares for is to 
maintain as guiding principles 
 only the NON-BROKEN
symmetries) 
to write a new set of operators for every new situation. This does not
necessarily contain all the operators of HPZH. This is one of the major 
shortcomings.

\subsubsection{\cviol and \pviol conserving $WWV$ couplings}
For the \cviol and \pviol part of the {\em HPZH} one writes
\beqn \label{pheno}
{\cal L}_{WWV}&=& -ie \left\{ \left[ A_\mu \left( W^{-\mu \nu} W^{+}_{\nu} - 
W^{+\mu \nu} W^{-}_{\nu} \right) \;+\;
\overbrace{ (1+\mbox{\boldmath $\Delta \kappa_\gamma$} )}^{\kappa_\gamma} 
F_{\mu \nu} W^{+\mu} W^{-\nu} \right] \right.
\nonumber\\
&+& \left. \mathrm{cotg}\;\theta_w \left[\overbrace{(1+ {\bf \Delta g_1^Z})}^{g_1^Z}
Z_\mu \left( W^{-\mu \nu} W^{+}_{\nu} - 
W^{+\mu \nu} W^{-}_{\nu} \right) \;
+ \;
\overbrace{(1+\mbox{\boldmath $\Delta \kappa_Z$} )}^{\kappa_Z} 
Z_{\mu \nu} W^{+\mu} W^{-\nu} \right] \right.
\nonumber\\
&+&\left. \frac{1}{M_{W}^{2}} 
\left( \mbox{\boldmath $\lambda_\gamma$} \;F^{\nu \lambda}+
\mbox{\boldmath $\lambda_Z$} \; 
\mathrm{cotg}\;\theta_w Z^{\nu \lambda}
\right) W^{+}_{\lambda \mu} W^{-\mu}_{\;\;\;\;\;\nu} \right\}
\eeqn

\noindent 
The first term, a photonic coupling, is not anomalous and is set by the
requirement that the $W$ kinetic term be $U(1)_{em}$ gauge invariant. 
This is the convection current. The $\kappa_\gamma$ term is a spin current. Its
coefficient may be anomalous as the magnetic moment of a composite particle 
can be anomalous, in the sense that its $g\neq 2$. Starting from the kinetic term and 
requiring only 
$U(1)_{em}$ will correspond to $\Delta \kappa_\gamma=-1$ 
(and $\lambda_\gamma=0$)~\cite{TDlee}. 
For those not working in the field and who want to get a feeling for what these 
form factors mean, suffice it to say that the combination 
$\mu_{W}=e/2M_W g_W=e (2+\Delta \kappa_\gamma + \lambda_\gamma)/2M_W$ describes 
the $W$ magnetic moment and $Q_{W}= - e (1+\Delta \kappa_\gamma
-\lambda_\gamma)/M_W^2$ its quadrupole moment \footnote{The deviations
from the \underline{minimal} gauge value are understood to be evaluated at
$k^2=0$.}. 
$(1+ {\bf \Delta g_1^Z})$ can be interpreted as the charge the ``$Z$ sees" in the
$W$. \\
\noindent 
Note that the $\lambda$ terms only involve the field strength, therefore they 
predominantly affect the production/interaction of transverse $W$'s, in other
words they do not usefully probe the \sb sector I am keen to talk
about here.  Pursuing this observation a little further one can easily
describe the distinctive effects the other terms have on different
reactions and the reason that some are found to be much better
constrained in some reactions than others. This is quite useful when
one tries to compare the limits LHC will set on these couplings as
compared to the high energy \epemt. 
\begin{figure*}[hbp]
\caption{\label{lkgfig}{\em The effect of the phenomenological
parameters on the vector boson pair production.}}
\begin{center}
\mbox{\epsfxsize=140mm\epsfysize=50mm\epsffile{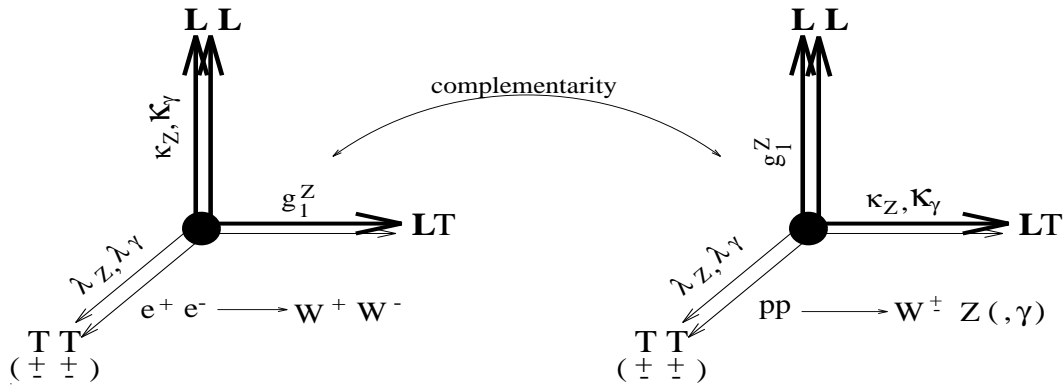}}
\vspace*{-1cm}
\end{center}
\end{figure*}
\vspace*{0.5cm}
%[0 0 829 342]
%
First, wherever you look, the $\lambda$'s live in a
world on their own, in the ``transverse world". If their effect is found to
increase dramatically with energy this is due to the fact that these
are higher order in the energy expansion (many-derivative operators). The other 
couplings can also grow with energy if a maximum number of longitudinals are
involved, the latter provide an enhanced strength due to the fact that 
the leading term of the longitudinal polarisation is $\propto \sqrt{s}/M_W$. 
This enhanced strength does not originate from the field strength! 
For instance, in \eewwt, $g_1^Z$ produces one $W$ longitudinal 
and one transverse: since the produced $W$ come, one from the field strength 
the other from the ``4-potential" (longitudinal) whereas the $\kappa$ terms 
produce two longitudinals and will therefore be better constrained in \eewwt. 
The situation is reversed in the case of $pp \ra WZ$. This also tells 
us how one may disentangle between different origins, the reconstruction of the
$W$ and $Z$ 
polarisation is crucial. I have illustrated this in 
fig.~\ref{lkgfig}, where I have reserved the {\bf thick} arrows for 
the ``important' directions. 

\noi Coming back to the warning
about the use of this phenomenological parameterisation outside its context, 
for instance to vector boson scattering. Even at tree-level it 
should be modified/extended to include appropriate accompanying 
``anomalous" quartic couplings. This is especially acute 
for $\lambda$ and $g_1^Z$, to restore 
$U(1)_{em}$ gauge invariance, at least...

\subsubsection{\cpviol preserving but \pviol violating operators}
\noi The inclusion of the other operators assumes violation of \cviol and/or 
\pviol. These may be searched for only if one reaches excellent 
statistics. Therefore the next 
operator which may be added is the \cpviol conserving but 
\pviol-violating Z coupling. 
In the {\em HPZH} parameterisation~\cite{HPZH} 
this coupling is introduced through $g_5^Z$: 
$\cl_2=- e g_5^Z c_w/s_w {\cal{O}}_2$
\beqn
{\cal{O}}_2=\epsilon^{\mu \nu \rho \sigma} 
\left(W_\mu^+ (\partial_\rho W_\nu) - (\partial_\rho W_{\mu}^{+}) W_\nu \right)
 Z_\sigma 
\eeqn

\section{Present Direct limits and the LEP data}
\begin{figure*}[htb]
\begin{center}
\caption{\label{d0ww.fig}{\em Present limits on $\Delta \kappa$ 
(with $\Delta \kappa_\gamma=\Delta \kappa_Z$) and $\lambda$
(with $\lambda_\gamma=\lambda_Z$) from the Tevatron. The 
$\star$ indicates a value meaning no spin-current for the 
$W$, it also leads to $Q_W$=0.}}
\mbox{\epsfysize=90mm\epsffile[143 179 471 490]{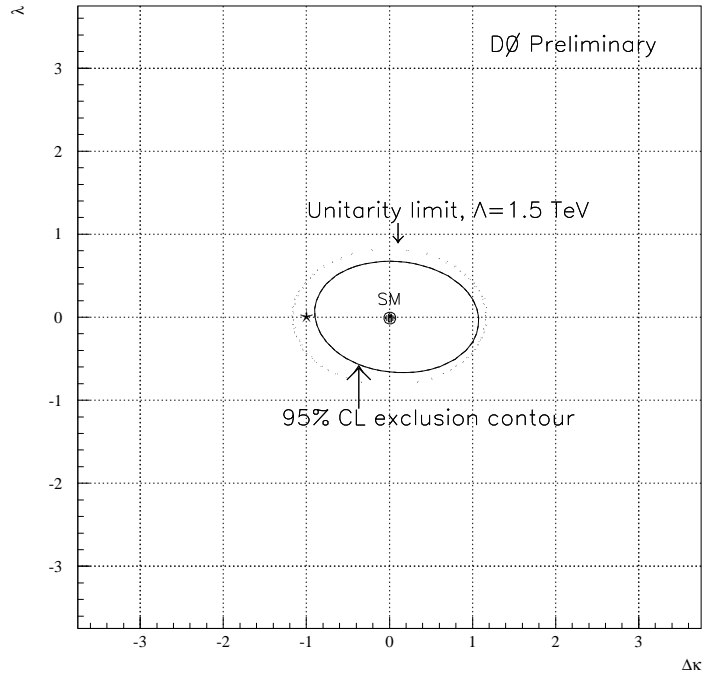}}
\end{center}
\end{figure*}
%[116 175 490 512]
%
There are some limits on the \cviol and \pviol couplings from
CDF/D0~\cite{CDFD0W95} extracted from the study of $WZ, WW$ and
$W\gamma$ production. From $W\gamma$ production D0 excludes, at
$80\%$CL, values that would correspond to only the convection current,
{\it i.e.} 
$\Delta \kappa_\gamma=-1, \lambda_\gamma=0$ are excluded. From a study 
of $WW$ and $WZ$ in CDF the existence of the $WWZ$ coupling is established at 
$95\%$CL ($g_1^Z \neq 0, \kappa_Z \neq 0$ is excluded). 
Thus one has with the latest results some {\em direct} confirmation that the 
weak bosons are self-interacting. However, the limits are still weak. With 
$\int {\cal L}= 13.8pb^{-1}$ of D0 data, 
 $-1.6 < \Delta \kappa_\gamma <1.8 (\lambda_\gamma=0)
 \;;\; -0.6< \lambda_\gamma < 0.6\;$. 
While a constrained global fit with $\lambda_\gamma=\lambda_Z$, 
$\kappa_\gamma=\kappa_Z (g_1^Z=1)$ gives 
$-0.89 < \Delta \kappa_V <1.07 \;;\;-0.66 < \lambda_V < 0.67$ based on 
$WW$ and $WZ$ events. 
%(which added up together amount to a predicted sample 
%of only $2.9 \pm 0.5$ events).\\

\noi 
The present Tevatron limits are hardly better than what we 
would extract 
from unitarity considerations as shown in Fig.~\ref{d0ww.fig}. The unitarity limit 
gives an order of magnitude for the upper values that these couplings have 
to satisfy if the scale of New Physics associated with these anomalous contributions 
is at 1.5 TeV. Nevertheless, as indicated by the star in 
Fig.~\ref{d0ww.fig}, there is direct empirical evidence that 
the $W$'s are self-interacting. 

\noi 
This said, I would like to argue that the present values are too large to be 
meaningful. These
are too large in the sense that they can hardly be considered as precision
measurements, a far cry from the precision that one has obtained on the vector-fermion 
couplings at LEP1! In the case of the Tevatron and $W$
self-couplings one is talking about deviations of order $100\%$! 
No wonder also that there is no entry for $(g-2)_W$ in the
PDB~\cite{pdb} which is even more than 
far cry when compared to $(g-2)_\mu$! Even the edm, electric dipole moment, 
of the $\tau$ (an unstable particle) is listed~\cite{pdb}. 

\begin{figure*}[htb]
\caption{\label{wloops.fig}{\em The need for $W$ loops from the present 
data (taken from~\protect\cite{Schildknecht95}). 
See the text for full explanation and comment on this plot.}}
\begin{center}
\mbox{\epsfxsize=150mm\epsfysize=105mm\epsffile[15 138 584 742]{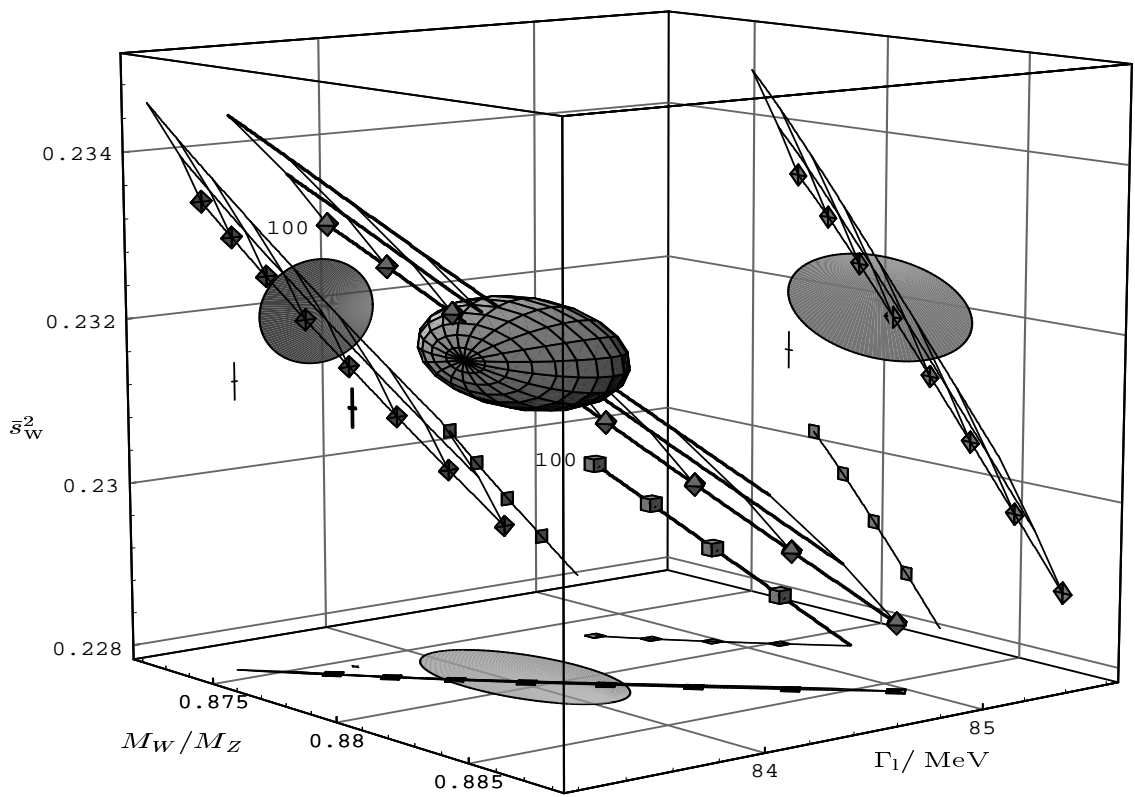}}
\end{center}
\end{figure*}

\noi In fact the LEP1 data have become so extraordinarily precise 
that they even very much influence our thinking and approach about the genuine 
non-Abelian structure of the \sm and the issue of gauge invariance. 
First, as pointed out 
already with the data available in 1994 by Gambino and Sirlin~\cite{Gambino} and 
Schildknecht and co-workers~\cite{Schildknecht94}, one is now sensitive to the 
genuine non-Abelian radiative corrections, and therefore to the presence of 
the tri-linear (and quadrilinear) couplings. The fermionic loops alone 
are no longer enough to reproduce the data 
\footnote{The separation of the bosonic one-loop 
contribution is meaningful in the sense of being gauge 
invariant~\cite{Sirlin80}.}. 
This important conclusion is reached even when restricting the analysis to the leptonic 
observables of LEP1 ( and in the case of~\cite{Schildknecht94,Schildknecht95} also to 
$M_W/M_Z$ from the UA2+CDF data) thus 
 dispelling any possible ambiguity 
that may enter through the hadronic observables (especially $R_b$). Hadronic uncertainties
only enter through the use of $\alpha(M_Z^2)$. To give an idea of how much needed these 
bosonic corrections are, Gambino and Sirlin~\cite{Gambino} have found that, 
when fixing the mass of the top in the range within the 
CDF/D0 measurements ($m_t=180$ GeV), there is a 
$7\sigma$ discrepancy between the predicted value of 
$\sin^2\theta_W(M_Z)|_{\overline{MS}}$ 
(amputated of its bosonic contribution) and that extracted from the data. 
$\sin^2\theta_W(M_Z)|_{\overline{MS}}$ is 
directly related to effective $\sin^2\theta_W^{eff}\equiv \bar s_W^2$ that expresses the leptonic 
asymmetries at the $Z$ peak. 
I can not resist showing the beautiful updated analysis (with the 1995 data) 
by Dittmaier, Schildknecht and Weiglein~\cite{Schildknecht95} as summarised in their plot 
in the three-dimensional space ($M_W/M_Z, \bar{s}^2_W, 
\Gamma_l$), see Fig.~\ref{wloops.fig}.

%[0 0 596 842]

The ``ball" is the $68\%$CL of the experimental 
data. The ``net" that 
the ball hits is the full \sm prediction and its array accounts for variations of 
$m_t$ and $M_H$. The $m_t$ line is depicted by diamonds starting with
$m_t=100$ GeV 
and increasing by steps of 20 GeV towards the lower plane . This line corresponds to 
a fixed Higgs mass of 100 GeV. The other lines are for $M_H=300$ GeV
and 1 TeV 
going from left to right. Keeping only the purely fermionic contribution gives the
line with cubes again starting with $m_t=100$ GeV and increasing by steps of 
20 GeV. It is clear that with $m_t\sim 180$ GeV the bosonic loops are mandatory. 
The thick cross represents the $\alpha(M_Z^2)$-Born approximation that certainly 
also no longer works as an approximation, as it used to~\cite{Novikov}. The projections into the corresponding 
two-dimensional plane which correspond to $83\%$CL are also shown. I think that the 
message 
is clear, it must be interpreted as an overwhelming support for the confirmation of the non-Abelian gauge 
structure of the \sm. It is important to stress, as can be seen from the plots, that 
one can not infer much on the presence of the Higgs, let alone its mass. At the moment its 
contribution, that only enters as $\log(M_H)$, can not be resolved experimentally. This 
dependence could also be interpreted as a cut-off dependence if the theory were 
formulated without the Higgs~\cite{Appelquist}. 
It is also amusing to note that there is an ambiguity
with interpreting the data as prefering a scenario with or without a Higgs, since one should 
specify precisely how one subtracts the Higgs contribution, not to mention that setting 
a high value for the Higgs mass means that one has lost control on the perturbative 
expansion. In the analysis by Gambino and Sirlin~\cite{Gambino}, this contribution 
is subtracted {\it \`a la} $\overline{MS}$. The effect of the subtraction is equivalent 
to calculating within the \sm with $M_H=113$ GeV. Thus, there is very little insight 
on the Higgs. This does not mean, on the other hand, that one has learnt absolutely 
nothing about symmetry breaking. It is now notoriously evident that some erstwhile 
popular (somehow natural) manifestations of technicolour do not 
stand the tests (see below). \\
%Of course this has not discouraged all 
% In fact some realisation of symmetry breaking 
%like technicolour are disfavoured. This does not mean that the idea of technicolour 
%has been killed some afficionadoes take the anomlous $R_b$ as strong evidence 
%for more sophisticated versions of technicolour that fits all present data 
%better than the \sm~\cite{Chivukula95}. \\
This remark about the sensitivity of the present data 
on the Higgs is in order since 
whether the Higgs is light or heavy or absent has an incidence on the physics 
of $W$'s at the future colliders. Meanwhile, one should be 
extremely cautious when trying to interpret the results of some latest 
global fits (not just the leptonic and the $M_W/M_Z$ that were discussed 
previously) as indicating a preference for a 
light Higgs, the subtraction issue being only one aspect. 
As stressed by 
Langacker~\cite{Langacker95} in his latest global analysis of the precision 
data, ``{\it the preference for small $M_H$ is driven almost entirely by $R_b$ and 
$A_{LR}$ both of which differ significantly from the \sm predictions. If these are due to 
large statistical fluctuations or to some new physics then the constraints on $M_H$ 
would essentially disappear.}" As an illustration of this point, it is amusing 
to note that 
the discrepancy in $R_b$ has triggered an interest in technicolour rather than 
killed it. It has been argued that some 
unconventional (non-commuting)
technicolour scenarios for symmetry breaking~\cite{Chivukula95} may explain all existing 
data better than the \sm. \\
As a conclusion about the impact of the low energy data on guiding us towards models of 
self-interacting $W$'s one should say that one should take the gauge invariance principle as 
sacrosanct but one should be open minded about the existence or the lightness of the 
Higgs. 

\section{The Hierarchy of couplings and their sizes}
In view of the verification of the \sm at better than the per-cent level 
(if one leaves aside the issue of $R_b$) and the confirmation of its inner 
workings at the quantum level, one should accept the fact that the model is a very 
good description of physics. Even though the fundamental issue of the Higgs is 
not resolved. One must also accept the principle of gauge invariance. But then how can one
justify allowing the operators in Eq.~5 (that we argued describe residual effects of 
New Physics?) and which are obviously not gauge invariant under the full 
\su local symmetry. 

\noi It is worth stressing again, contrary to the fierce attack~\cite{Ruj}
that blames the above HPZH Lagrangian (Eq.~\ref{pheno}) for not being locally
gauge invariant and leading to trouble at the quantum level, that as
the lengthy introduction has shown all the above operators can be made
gauge invariant, by unravelling and making explicit the compensating
Goldstone fields and extra vertices that go with the above. 
As is evident with the mass terms, the term in Eq.~\ref{wmass} is not 
gauge invariant but this is only because it has been truncated from 
its larger part, Eq.~\ref{higgspart} or Eq.~\ref{masscov}. 
Under this light, the HPZH parametrisation
should be considered as being written in a specific gauge and that after this 
gauge (unitary) has been chosen it is non-sensical to speak of gauge
invariance~\cite{Cliff,Hawai}. 
But of course, it is much much better to keep the full symmetry so that one can
apply the Lagrangian to any situation and in any frame. There is another
benefit in doing so. If the scale of new physics is far enough compared to the
typical energy where the experiment is being carried out\footnote{If this is
not the case then we should see new particles or at least detect their tails.},
then one should only include the first operators in the energy expansion, beyond
those of the \sm. Doing so will maintain some constraints on the parameters
$\lambda, g_1^Z, \Delta \kappa$. These constraints will of course be lost if you
allow higher and higher order operators or allow strong breaking of custodial
symmetry, in both cases rendering the situation chaotic while LEP1 shows and
incredible regularity. It is highly improbable that the order and symmetry is
perturbed so badly. 

So what are these operators that describe the self-couplings when one restricts 
one-self to next-to-leading operators by exploiting the \su and the custodial 
symmetry? and how are they mapped on the HPZH phenomenological 
parameters? The leading operators are of course those that describe the 
minimal standard model(Eqs.~\ref{higgspart},~\ref{kinetic} or it minimal Higgs-less 
version Eqs.~\ref{higgspart},~\ref{kinetic}). 
These are given in Table~\ref{linearvsnonlinear} for the linear~\cite{BuchWy,Ruj} as well as the
non-linear realisation~\cite{Hawai,Holdom} to
bring out some distinctive features about the two approaches:
%With these few points spelled out, we arrive at the most {\em probable} 
%set of yet-untested operators, within a linear~\cite{BuchWy,Ruj} or a 
%non-linear~\cite{AppelquistLong,Holdom,Espriu,FLS,BDV,Feruglio,AppelquistWu} 
%realisation of \ssb. 
\begin{table*}[hbtp]
%\begin{table*}[h]
\caption{\label{linearvsnonlinear}
{\em The Next-to-leading Operators describing the $W$ Self-Interactions which 
do not contribute to the $2$-point function.}}
\vspace*{0.3cm}
\centering
\begin{tabular}{|l||l|}
\hline
{\bf Linear Realisation \hspace*{0.1cm}, \hspace*{0.1cm} Light Higgs}&
{\bf Non Linear-Realisation \hspace*{0.1cm}, \hspace*{0.1cm} No Higgs}\\
\hline
&\\
${{\cal L}}_{B}=i g' \frac{\epsilon_B}{\Lambda^2} (\cd_{\mu}
\Phi)^{\dagger} B^{\mu \nu} \cd_{\nu} \Phi$&
${{\cal L}}_{9R}=-i g' \frac{L_{9R}}{16 \pi^2} \tr ( \B^{\mu \nu}\cd_{\mu}
\Sigma^{\dagger} \cd_{\nu} \Sigma )$ \\
&\\
${{\cal L}}_{W}=i g \frac{\epsilon_w}{\Lambda^2} (\cd_{\mu}
\Phi)^{\dagger} (2 \times \W^{\mu \nu}) (\cd_{\nu} \Phi)$&
${{\cal L}}_{9L}=-i g \frac{L_{9L}}{16 \pi^2} \tr ( \W^{\mu \nu}\cd_{\mu}
\Sigma \cd_{\nu} \Sigma^{\dagger} ) $ \\
&\\
$\cl_{\lambda} = \frac{2 i}{3} \frac{L_\lambda}{ \Lambda^2} 
g^3 \tr ( \W_{\mu \nu} \W^{\nu \rho} \W^{\mu}_{\;\;\rho})$&
$\;\;\;\;\;\;\;\;---------\;\;\;\;$\\
&\\
$\;\;\;\;\;\;\;\;---------\;\;\;\;$&
$\cl_{1}=\frac{L_1}{16 \pi^2} \left( \tr (D^\mu \Sigma^\dagger D_\mu \Sigma) 
\right)^2\equiv \frac{L_1}{16 \pi^2} {{\cal O}}_1$ \\
$\;\;\;\;\;\;\;\;---------\;\;\;\;$&$
\cl_{2}=\frac{L_2}{16 \pi^2} \left( \tr (D^\mu \Sigma^\dagger D_\nu \Sigma)
\right)^2 \equiv \frac{L_2}{16 \pi^2} {{\cal O}}_2$ \\
& \\
\hline
\end{tabular}
\end{table*}

\noi By going to the 
physical gauge, one recovers the phenomenological parameters with the 
{\em constraints}:
\beqn \label{constraints}
\Delta\kappa_\gamma&=&
\frac{e^2}{s_w^2} \frac{v^2}{4 \Lambda^2}
(\epsilon_W+\epsilon_B )=
\frac{e^2}{s_w^2} \frac{1}{32 \pi^2} \left( L_{9L}+L_{9R} \right)
\nonumber \\
\Delta\kappa_Z&=&\frac{e^2}{s_w^2} \frac{v^2}{4 \Lambda^2}
(\epsilon_W -\frac{s_w^2}{c_w^2} \epsilon_B) =
\frac{e^2}{s_w^2} \frac{1}{32 \pi^2} \left( L_{9L} 
-\frac{s_w^2}{c_w^2} L_{9R} \right) 
\nonumber \\
\Delta g_1^Z&=&\frac{e^2}{s_w^2} \frac{v^2}{4 \Lambda^2}
(\frac{\epsilon_W}{c_w^2})=
\frac{e^2}{s_w^2} \frac{1}{32 \pi^2} \left(\frac{ L_{9L}}{c_w^2} \right)
\nonumber \\
\lambda_\gamma&=&\lambda_Z=\left(\frac{e^2}{s_w^2}\right)
L_\lambda \frac{M_W^2}{\Lambda^2}
\eeqn

\subsection{Quartic couplings}
Note that $L_{9L,W,\lambda}$ do give quadri-linear bits but
these are imposed by gauge invariance. This is confirmed by many analyses. 
%set of yet-untested operators, within a linear~\cite{BuchWy,Ruj} or a 
%non-linear~\cite{AppelquistLong,Holdom,Espriu,FLS,BDV,Feruglio,AppelquistWu} 
%realisation of \ssb. 
Going to the physical gauge, the quartic couplings from the chiral 
approach are 
\beqn
\cl_{WWV_{1}V_{2}} &=& -e^2 \left\{ \left(
A_\mu A^\mu W^{+}_{\nu} W^{- \nu} - A^\mu A^\nu W^{+}_{\mu} 
W^{-}_{\nu} \right) \right. \nonumber \\
&+& 2 \frac{c_w}{s_w} (1+\frac{l_{9l}}{c_w^2}) \left(
A_\mu Z^\mu W^{+}_{\nu}W^{-\nu} - \frac{1}{2} 
A^\mu Z^\nu ( W^{+}_{\mu}W^{-}_{\nu} + W^{+}_{\nu}W^{-}_{\mu} ) \right) 
\nonumber \\
&+&\frac{c_w^2}{s_w^2} (1+\frac{2 l_{9l}}{c_w^2}-\frac{l_-}{c_w^4}) \left(
Z_\mu Z^\mu W^{+}_{\nu}W^{-\nu} - Z^\mu Z^\nu W^{+}_{\mu}W^{-}_{\nu} \right) 
\nonumber \\
&+& \frac{1}{2 s_w^2} (1+2 l_{9l}-l_-) 
 \left(W^{+\mu} W^{-}_{\mu} W^{+\nu} W^{-}_{\nu} -
W^{+ \mu} W^{+}_{\mu} W^{-\nu}W^{-}_{\nu} \right) \nonumber \\
&-&\frac{l_+}{2s_w^2}\left( \left( 3 W^{+\mu} W^{-}_{\mu} W^{+\nu} W^{-}_{\nu}+
W^{+ \mu} W^{+}_{\mu} W^{-\nu}W^{-}_{\nu} \right) \right. \nonumber \\
&+&\frac{2}{c_w^2} \left. \left. \left(Z_\mu Z^\mu W^{+}_{\nu}W^{-\nu} + Z^\mu Z^\nu W^{+}_{\mu}W^{-}_{\nu}
 \right) + \frac{1}{c_w^4} Z_\mu Z^\mu Z_\nu Z^\nu \right)\right\} \nonumber \\
&\;& {\rm with}\;\;\;\;\;\;l_{9l}=\frac{e^2}{32 \pi^2 s_w^2} L_{9L}\;\;\; ;\;\;\;l_\pm=\frac{e^2}{32 \pi^2 s_w^2}
(L_1 \pm L_2)
\eeqn
Note that the genuine trilinear $L_{9L}$ gives structures analogous to the \sm. 
The two photon couplings (at this order) are untouched by anomalies.
The $L_{1,2}$, which have no equivalent at this order in the linear 
approach, 
contribute only to the quartic couplings of the massive states and thus 
genuinely describe quartic couplings.

\subsection{\em Catch 22:}
Aren't there other operators with the same symmetries that appear at the {\em same
level} in the hierarchy and would therefore be as likely? \\
\noi \underline{Answer:} YES. And this is an upsetting conceptual problem. 
On the basis of the above symmetries, one can not help it, but 
there are other operators which contribute to 
the tri-linear couplings and have a part which corresponds to 
bi-linear anomalous 
$W$ self-couplings. Because of the latter and of the unsurpassed precision 
of LEP1, these operators are already very much {\em unambiguously} 
constrained at the same level that operators describing a breaking 
of the global $SU(2)$ symmetry (especially after the slight breaking due 
to the $t-b$ mass shift has been taken into account). Examples of
such annoying operators in the two approaches are
\beqn
{{\cal L}}_{WB}&=&g g' \frac{\epsilon_{WB}}{\Lambda^2}\; \left(
\Phi^{\dagger} \times \W^{\mu \nu} \Phi \right) B_{\mu \nu} 
\nonumber \\
{{\cal L}}_{10}&=&g g' \frac{L_{10}}{16 \pi^2} \tr ( \B^{\mu \nu}
\Sigma^{\dagger} \W^{\mu \nu} \Sigma ) \longrightarrow L_{10}=-\pi S\simeq \frac{4 \pi
s_W}{\alpha} \epsilon_3 
\eeqn

\begin{figure*}[htbp]
\caption{\label{Langacker.fig}{\em Present limits on 
\protect$L_{10}$ and \protect$T_{new}$. 
From~\protect\cite{Langacker95}. 
The projection of the \protect$L_{10}$ constraint from the fit is also
shown (``bar").}}
\mbox{%
\epsfxsize=\textwidth\epsfysize=205mm\epsffile[113 321 521 753]{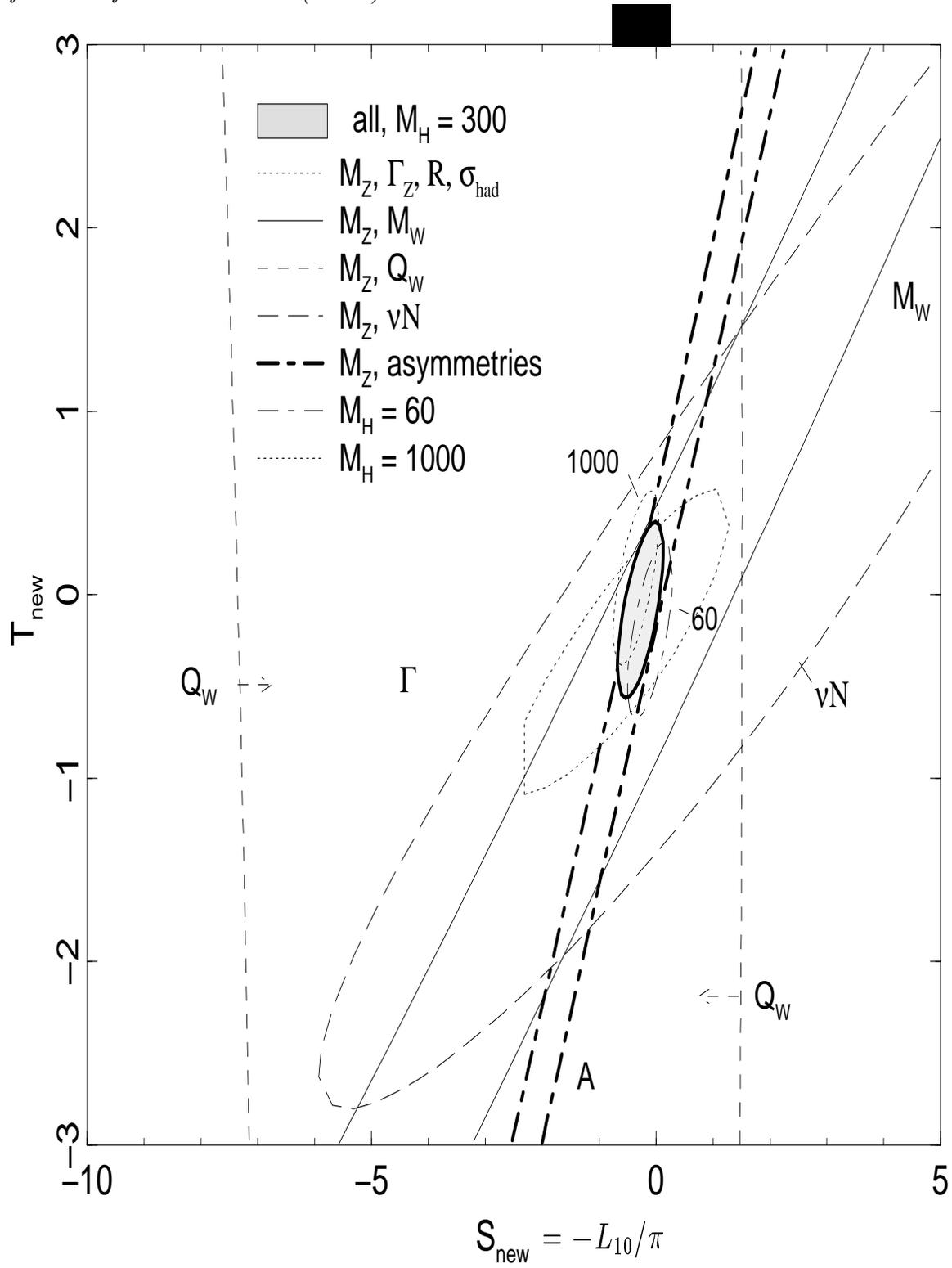}}
\end{figure*}

Limits on the coefficient $L_{10}$ from a global fit to the existing data are shown in the 
figure borrowed from~\cite{Langacker95}, see Fig.~\ref{Langacker.fig}. Allowance for a 
possible deviation of the $\rho$ parameter from unity (slight breaking of the global 
symmetry due to physics beyond the \sm) is made: $\rho=1+\alpha T_{new}$. 
%[20 300 596 842]
%[113 321 521 744]
One can see that current limits indicate that $-0.7<L_{10}<2.4$ 
for a {\bf 2}-parameter fit. 
This is really small, so small that if the other $L_i$'s, say, were of
this order it would be extremely difficult to see an effect at the
next colliders. So why should the still not-yet-tested operators be
larger and/or provide more stringent tests? This is the naturalness
argument which, in my view, is {\em the} essential point of~\cite{Ruj}. 
One can try hard to find models with no contributions to $L_{10}$. One
way out is to associate the smallness of $L_{10}$ to a symmetry that
forbids its appearance, in the same way that the custodial $SU(2)$
symmetry prevents $\Delta \rho$ or leads to so small values of
$T_{new}$. 
$L_{10}$ represents the breaking of the axial $SU(2)$ global
symmetr~\cite{InamiL10}. But the construction of models is
difficult. Models that make do with the $L_{10}$ constraint include
some dynamical vector models that deviate from the usual scaled up
versions of QCD by having (heavy) axial vectors like the extended BESS
model~\cite{BESS}. 
The latter implements an $(SU(2)_L \times SU(2)_R)^3$ symmetry, as
discussed by D.~Dominici in this volume. Axial vectors are included in
the analysis presented in the talk of Tanabashi~\cite{Tanabashi}. 

For the light Higgs scenario one has decoupling of the heavy particles
so that the predicted $\epsilon_{WB}$ is small and does not have much
incidence on the LEP1 data. A case in point is the minimal SUSY
contribution. But at the same time one expects the other $\epsilon_i$
to be of the same order. The new physics in the light Higgs scenario
may start to have an impact only if one is close to the threshold of
production of the new particles which circulate in the loops. But then
the effective approach through anomalous effective couplings is
clearly not applicable. For instance, in \epemwwt one can not just
subtract the parts that map into $\Delta g_1^Z$ etc... but one should
take the full 1-loop contribution to the process, boxes for instance
would not be negligible in these configurations. These expectations
are borne out by explicit calculations~\cite{Masiero}. One can, of
course, hope to improve on the limits from LEP1, by exploiting the
$k^2$ enhancement factor from these operators, in 2 fermion production
as has been discussed by R.~Szalapski in the parallel session~\cite{Szalapski}.

To continue with my talk I will assume that $L_{10}\sim 0$ can be
neglected compared to the other operators. This said, I will not
completely ignore this limit and the message that LEP1 is giving us,
$L_{10}\sim 1$, especially that various arguments about the ``natural
order of magnitude"~\cite{Wudkanatural} for these operators should
force one to consider a limit extracted from future experiment to be
meaningful if $|L_i|<\sim 10$. This translates into $\Delta\kappa,
\dgz <\sim 10^{-2}$. Note that the present Tevatron limits if they
were to be written in terms of $L_{9}$ give $L_9\sim 10^3$!!! 

\noi With this caveat about $L_{10}$ and the like, let us see how the 2 opposite 
assumptions about the lightness of the Higgs differ in their most probable
effect on the $W$ self-couplings. First the tri-linear couplings $\lambda$ is
relegated to higher orders in the heavy Higgs limit(less likely). 
This is as expected: transverse
modes are not really an issue here. The main difference is that with a heavy
Higgs, genuine quartic couplings contained in $L_{1,2}$ are as likely as the tri-linear
 and, in fact, when
contributing to $WW$ scattering their effect will by far exceed that of the tri-linear.
This is because $L_{1,2}$ involve essentially longitudinals. This is another way
of arguing that either the Higgs exists or 
one should expect to ``see something" in $WW$
scattering. Note also that $L_{9R, B}$ is not expected
to contribute significantly in $pp \ra WZ$ since it has no contribution to
$\Delta g_1^Z$ (see Fig.~\ref{lkgfig}). This is confirmed by many analyses. 

%set of yet-untested operators, within a linear~\cite{BuchWy,Ruj} or a 
%non-linear~\cite{AppelquistLong,Holdom,Espriu,FLS,BDV,Feruglio,AppelquistWu} 
%realisation of \ssb. 

\section{Future Experimental Tests}
With the order of magnitude on the $L_i$ that I have set as a meaningful 
benchmark,
one should realise that to extract such
(likely) small numbers one needs to know the \sm cross sections with a precision
of the order of $1\%$ or better. From a theoretical point of view 
this calls for the need to include the radiative
corrections especially the initial state radiation. 
Moreover one should try to
extract as much information from the $W$ and $Z$ samples: reconstruct the
helicities, the angular distributions and correlations of the decay products.
These criteria mean precision measurements and therefore we expect \epemt
machines to have a clear advantage assuming that they have enough energy.
Nonetheless, it is instructive to refer to Fig.~\ref{lkgfig} to see
that $pp$ machines could be complementary. 

In the following, one should keep in mind that all the extracted
limits fall well within the unitarity limits. I only discuss the
description in terms of ``anomalous couplings" below an effective cms
energy of a VV system 
$\sim 4 \pi v \sim 3-4$ TeV, without the inclusions of resonances%
\footnote{At this conference, this has been discussed by Tim Barklow.}.  
Moreover, I will not discuss the situation when parameters are dressed with 
energy dependent form factors or any other scheme of unitarisation that
introduces more model dependence on the extraction of the limits. 
These limits are given in terms of the chiral lagrangian parameters
$L_{9L,R}$ or equivalently using Eq.~\ref{constraints} in terms of $L_{B,W}$. They can also be 
re-interpreted in terms of the more usual $\kappa_V,g_1^Z$ with the constraint given
by Eq.~\ref{constraints}, in which case the $L_{9L}$ axis is directly proportional
to \dgzt. The reason for prefering to give the limits in the parameterisation $L_{9L}-L_{9R}$ 
is that, as argued above, by the time the NLC is built and run a 
few days one will know whether the Higgs is there or 
not, if it had not been discovered before. 
If it is around, the most interesting physics will be that of the Higgs and 
most likely supersymmetry. It is unlikely that one would learn more about symmetry breaking 
by searching for anomalous couplings in the tri-linear and quadri-linear couplings (that would have 
to be implemented through the linear approach) than by probing the supersymmetric spectrum or
critically probing the properties of the Higgs~\cite{Gounarishiggs}. 

\subsection{Tri-linear couplings}
\begin{figure}[bth]
\begin{center}
\mbox{\epsfysize=130mm\epsffile{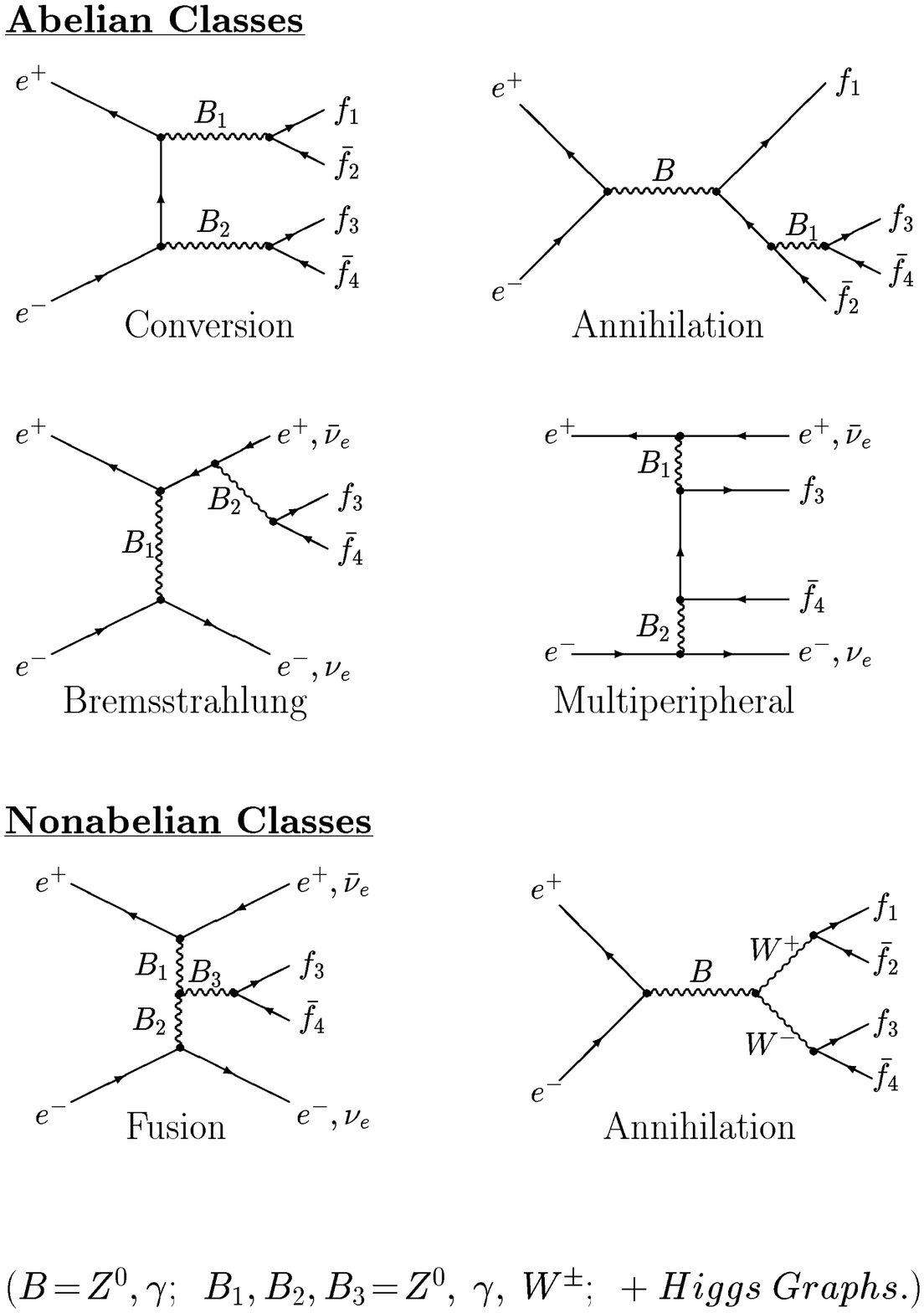}}
\caption{\it Four-fermion production classes of diagrams from~\protect\cite{TRieman}.}
\label{Lehner.fig}
\end{center}
\end{figure}

In principle all $W/Z$ production processes could be used to search for possible anomalies
in the self-couplings. As seen from Fig.~\ref{wfactory} there are quite a few mechanisms. 
It is however clear that the best are those that have the largest 
cross sections and that are cleanest, especially after implementing the acceptance factors. 
Some of the largest cross sections are mostly due to forward events due to t-channel 
structures 
that are mostly dominated 
by transverse states, and consequently are less interesting. 
In this respect 
\epemwwt is one of the most promising and important cross sections. 
The full radiative corrections have been computed 
by two different 
groups and the results are found to agree perfectly~\cite{eewwrc}. 
These corrections are quite large and are due essentially to initial
state radiation (ISR) in particular from the hard collinear
photons. The latter drastically affect some of the distributions
which, at tree-level, seem to be good New Physics discriminators
because they are responsible for the boost effect that redistributes
phase space: 
this leads to the migration of the forward $W$ (dominated by the t-channel 
neutrino contribution)
into the backward
region and results in a large correction in the backward region. Precisely the
region where one would have hoped to see any s-channel effect more clearly.
Second, if one reconstructs the polarisation of the $W$ 
without taking into account the energy loss, one may ``mistag" 
a transverse $W$ for a longitudinal, thereby introducing a huge correction in the small
tree-level longitudinal cross sections, which again is 
 particularly sensitive to New Physics. Since at the NLC one has to take into account not only 
the bremsstrahlung but also the important beamstrahung contribution to the ISR, this 
contamination must be reduced to a minimum. Fortunately, there is a variety of cuts 
to achieve this, such as accolinearity cuts or keeping into the useful event samples only those 
events where the energies of the $W$'s add up to the cms energy. 
One welcome feature at the Linear Collider energies, compared to LEP2, 
is that $\cal O(\alpha^2)$ corrections are
negligible; moreover we do not have to worry about 
Coulomb singularities~\cite{Coulomb}. The latter enhance the
cross section of a pair of slowly moving 
particles which have opposite charges. As for the genuine weak corrections 
the upshot is that the use of $G_F$ and a running $\alpha_{em}$ 
absorbs a large part of the 
weak corrections~\cite{Wim}. As a results there is no $m_t^2$ dependence, however some mild log dependence 
on the heavy particles (notably the Higgs) may survive. The latter could be responsible 
for delaying unitarity to higher energies and from this point of view mimick the effect of 
anomalous couplings. \\
In actual life the $W$'s are only observed through their decay products and therefore the
$W$-pair production observables are in fact $4$-fermion observables. Unfortunately not all 
the 4-f observables are intimately related to the doubly resonant $WW$ production. Depending on the 
final state many other 4-f final states (especially with electrons) that do 
not proceed via $WW$
 are present, see Fig.~\ref{Lehner.fig}. 
The extra processes for $4$-f production could be potential background to $WW$ production and most 
importantly could mask the subtle effect that one would like to extract. Hence the need to know their cross
sections 
as precisely as possible and find ways to subtract them. This subtraction, for instance requiring
the invariant mass of both pairs of fermions to be consistent 
with the $W$ mass 
within the experimental resolution, can be done but at the expense of less statistics. Not that this is a problem
with $WW$. 
From a theoretical point of view one can formally always find a
 prescription for isolating 
the $WW$ process. One may also prefer to keep all final states especially that there 
are topologies 
which are not associated with $WW$ production but which nevertheless involve the tri-linear
couplings. Single-$W$ production as materialised through a fusion process 
(see Fig.~\ref{Lehner.fig}) 
is such a process. It is then worth implementing non-standard couplings to the complete 
set of diagrams that lead to $4$-f and see if the sensitivity is 
enhanced. Many programs that 
calculate 4-fermion final states have been developped recently, especially for 
LEP2~\cite{smplep2,evgenlep2}. These should
be optimised when moving to higher energies. Not all of them calculate all configurations of 
4f. Those that involve electrons in the final states are hardest to calculate especially if one
wants to keep the electrons exactly forward. In fact only a few manage to do this since this
requires to keep the electron mass explicitely. Moreover one has to be careful with the
implementation of the $W$ width. The problem of how to introduce the $W$ width is quite subtle 
and it is fair to say that at the moment though a few 
schemes have been suggested~\cite{kurihara95,Baurwidth,Argyres95} 
there is no definite answer as yet on how to handle the width. \\
So at the moment one has full radiative corrections to the doubly resonant 
$WW$ process (as if the $W$ were stable) 
%or for 
%the case of single $W$ production to the $e \gamma \ra W \nu$ process
 and full contributions to the 
$2\ra 4$ process at tree-level only. The best of both worlds is to have the radiatively corrected 
$4$ final states. This is a horrendous task. However it is good to know that what one expects to be
the largest corrections, ISR, can be easily implemented. Almost all of the 
 $4$-fermion codes
convolute with a radiator. Some codes like grc4f~\cite{grace} can
implement all ingredients: non-zero fermion masses, ISR, polarisation and anomalous couplings. 

For extracting, or putting limits, on the anomalous couplings some extensive 
studies that aim at 
getting all the information which is contained in the 4f final states have 
been performed. As we stressed all along, the most important part is the 
longitudinal part of the $W$, which can be reconstructed by looking at 
the angular distributions of the fermion that results from the decay the $W$. 
In fact because of the $V-A$ structure of the $W$ interaction to 
fermions, 
these angular distributions (both polar $\theta_i$ and azimuthal $\phi_i$, the latter 
requires tagging the charge of the fermion) constitute an excellent polarimeter. 
Ideally one
would like to have access to all the helicity amplitudes and the density 
matrix elements of the $WW$ process. Of course this neglects non resonant 
$WW$ diagrams or presumes the invariant mass cuts to be 
rather stringent such that one only 
ends 
up with the $WW$ process. Until very recently this has been the most widely used 
approach. In this approximation the theoretical issues are quite transparent.
One starts from the helicity amplitudes of the 
$e^+ e^-(\lambda) \ra W^-(\tau_1) W^+(\tau_2)$ where $\lambda, \tau_{1,2}$ are 
the respective helicities. ``Assuming" that one knows the velocity of the 
$W$ (or three momentum $|\vec{p}|$), which means that some ``kneading" has been 
done about controlling ISR, and also the decay angles of the charged fermion as 
measured in the rest frame of the parent $W$ ($\theta_i$,$\phi_i$ $i=1,2$), 
all the information is contained in the five-fold differential cross section
\beqn
\frac{ {\rm d}\sigma(\epemww\ra f_1 \bar f_2 f_3 \bar f_4)}
     { {\rm d}\cos \theta {\rm d}\cos \theta_1 {\rm d}\phi_1
      {\rm d}\cos \theta_2 {\rm d}\phi_2 }=Br^{f_1 \bar f_2} Br^{f_3 \bar f_4} 
\frac{1}{16\pi s}\frac{|\vec{p}|}{\sqrt{s}} \left( \frac{3}{8 \pi}\right)^2
\nonumber \\
\sum_{\lambda \tau_1 \tau_2 \tau'_1 \tau'_2}
 F_{\tau_1\tau_2}^\lambda (s,\cos \theta)\; F_{\tau'_1\tau'_2}^{*\lambda} (s,\cos \theta)
\; D_{\tau_1 \tau'_1} (\theta_1 ,\phi_1) \; D_{\tau_2 \tau'_2} (\pi-\theta_2, \phi_2+\pi)
\nonumber \\
\equiv 
\frac{{\rm d}\sigma(\epemww)}{{\rm d}\cos \theta}\left( \frac{3}{8 \pi}\right)^2 
\sum_{\tau_1 \tau_2 \tau'_1 \tau'_2} \rho_{\tau_1 \tau_2 \tau'_1 \tau'_2}\;
D_{\tau_1 \tau'_1} (\theta_1 ,\phi_1) \;
D_{\tau_2 \tau'_2} (\pi-\theta_2, \phi_2+\pi)
\eeqn
 
$\theta$ is the scattering angle of the $W$ and $\rho$ is the density matrix that 
can be projected out. Note that one separates the decay and the production parts. 
To be able to reconstruct the direction of the charged $W$ and to have least 
ambiguity 
the best channel is the semi-leptonic channel 
$W^\pm\ra l \nu_l, W^\mp\ra jj$ with $l=e,\mu$. 
Note however that if the charge of the jet is not reconstructed then we can not fully 
reconstruct the helicity of the $W$ that decays into jets, though an averaging may be 
done. It is also important to take into account initial electron polarisation, this 
in fact helps. For instance we can choose to concentrate on the $s$-channel only 
where the anomalous contribute, even if the statistics are lower. 
The BMT collaboration~\cite{BMT} has exploited the above approach where the fit has been 
done on the {\em reconstructed} density matrix elements and the limits on the parameters 
of the anomalous operators are extracted through a $\chi^2$ minimisation method. It is found~\cite{Sekulin}
that the limits are much improved if one fits directly on the $5$ angular variables 
and uses a maximum likelihood method. Barklow~\cite{Barklow}
 has also implemented this latter approach 
and looked also at the effect of ISR and beam polarisation at the linear collider. 
Preliminary investigations indicate that one can improve on the limits by also 
including the $4$-jets sample and perhaps the $\tau$ final 
state~\cite{Jbhansen}. \\
\noi Alternatively, one would like to fit on all the available variables of the complete
set of a particular $4$-f final state and not
restrict the analysis to the resonant diagrams. Excalibur~\cite{Excalibur}, 
Erato~\cite{Erato} and grc4f~\cite{grace} allow to do such an analysis even with the 
inclusion of ISR. grc4f can even study single $W$ production (no cut on the 
forward electron), for a preliminary analysis see~\cite{kuriharasingle}. A thorough 
investigation for the linear collider energies taking into account the final $W$ 
width (all diagrams) using the maximum likelihood fit method has been done 
by Gintner, Godfrey and Couture~\cite{GGC}. ISR is not implemented but their results which we 
will comment upon in a moment, are quite telling especially that one can compare them 
with those based on the resonant diagrams. 

\subsection{Other Processes at and other modes of the linear collider}
Perhaps the only disadvantage with the $WW$ process is that it is not easy to 
disentangle the $WW\gamma$ from the $WWZ$ couplings. To single out one of them 
one can exploit $\epem \ra \nu \nu_e \gamma$~\cite{eenng} or 
$\epem \ra \nu \nu_e Z$~\cite{eennz}. However, it should be pointed out 
that in specific models 
or in the chiral approach for example, the $WW\gamma$ and $WWZ$ couplings are related, 
see Eq.~\ref{constraints}. Taking these constraints into account, 
the latter reactions do not seem to be competitive 
with $WW$. One of the reasons, especially for $\nu \nu \gamma$, is that the 
helicity structure is not as rich as in $WW$: all final particles 
do not have ``interesting 
helicities". $WW\gamma$ and $WWZ$~\cite{nousee3v} production in \epemt
 are quite interesting but, as far as tri-linear 
couplings are concerned, they can compete with the $WW$ only for TeV energies since they 
suffer from much lower rates~\cite{nousee3v}. However they are very suited 
to study possible quartic couplings. In this respect $WWZ$ can probe the all 
important $L_{1,2}$. \\
In the \ememt mode one can investigate the tri-linear couplings through 
$e^-e^- \ra e^- W^-\nu$~\cite{CuypersenW} this may be considered as the equivalent \epemt 
reaction $\epem \ra \nu \nu_e Z$ and has the same shortcomings. Cuypers has investigated 
the potential of this reaction in probing the tri-linear couplings by taking into account 
the possibility of polarised beams, fits have only been done to the scattering 
angle of the the final electron. \\
The $\gamma \gamma$ mode is of course ideal for the photonic couplings especially that 
here the cross section for $WW$ production is
huge~\cite{nousggvvref}. 
In the chiral approach this reaction will only constrain the
combination $L_{9L}+L_{9R}$, but as we will see, in conjunction with
the \epemt mode this is quite helpful. To probe the $WWZ$ couplings in
\gamgamt one can use the rather large
$WWZ$~\cite{nousggwwz,brasilggwwz} rate. 
The corresponding reactions in the $e\gamma$ mode are \
$e \gamma \ra W \nu$~\cite{egnw} and 
 $e \gamma \ra W \nu Z$~\cite{egnwz}. As with the two-body 
reaction in \epemt (\epemwwt), the 
radiative corrections for $e \gamma \ra W \nu$~\cite{egwnrc} and 
$\gamma\gamma\ra W^+ W^-$~\cite{ggwwrc} have been calculated. 

\subsection{Limits from the LHC}
Before discussing the limits on the parameters of the chiral Lagrangian that one 
hopes to achieve at the different modes of the linear colliders it is essential 
to compare with the situation at the LHC. For this we assume the high luminosity 
option with 100 fb$^{-1}$. 
These limits are based on a very careful
study~\cite{Baur} that includes the very important effect of the QCD
NLO corrections as well as implementing the full spin correlations for the most
interesting channel $pp \ra WZ$. $WW$ production 
with $W\ra jets$ production is fraught with a huge QCD background, while the
leptonic mode is extremely difficult to reconstruct due to the 2 missing
neutrinos. Even so, a thorough investigation 
(including NLO QCD corrections) 
for this channel has been done~\cite{Baurnloww}, which confirms the 
superiority of the $WZ$ channel as we will see.
 The NLO corrections for $WZ$ production are huge, especially 
in precisely the regions
where the anomalous are expected to show up, for instance, high $p_T^Z$. 
In the inclusive cross section this is mainly due to, first, the importance 
of the subprocess $q_1 g\ra Z q_1$ (large gluon density at the LHC) followed by 
the ``splitting" of the quark $q_1$ into $W$. The probability for this splitting
increases with the $p_T$ of the quark (or Z): Prob$(q_1\ra q_2 W) \sim
\alpha_w/4\pi ln^2(p_T^2/M_w^2)$. To reduce this effect one has to define 
an exclusive cross section that should be as close to the LO $WZ$ cross section 
as possible by cutting on the extra high $p_T$ quark (dismiss any jet with 
 $p_T^{{\rm jet}}>50\ GeV, |\eta_{{\rm jet}}|<3$). This defines a NLO 
$WZ +``0{{\rm jet"}}$ cross section which
is stable against variations in the choice of the $Q^2$ but which nonetheless can
be off by as much as $20\%$ from the prediction of Born \sm result. 

\section{Comparisons and Discussion}
\begin{figure*}[htbp]
\caption{\label{l9fig}{\em Comparison between the expected bounds on
the two-parameter space $(L_{9L},L_{9R}) \equiv (L_W,L_B) \equiv
(\Delta g_1^Z, \Delta \kappa_\gamma)$ (see text for the conversions)
at the NLC500 (with no initial polarisation) and LEP2. 
The NLC bounds are from $e^+e^- \ra W^+W^-\;,W^+W^-\gamma, W^+W^-Z$
(for the latter these are one-parameter fits), 
$\gamma\gamma\ra W^+W^-$ and $e^-e^-\ra W^- \nu e^-$. 
Limits from a single parameter fit are also shown (``bars").}} 
\begin{center}
\mbox{\epsfxsize=\textwidth\epsfysize=190mm\epsffile[25 108 557 679]{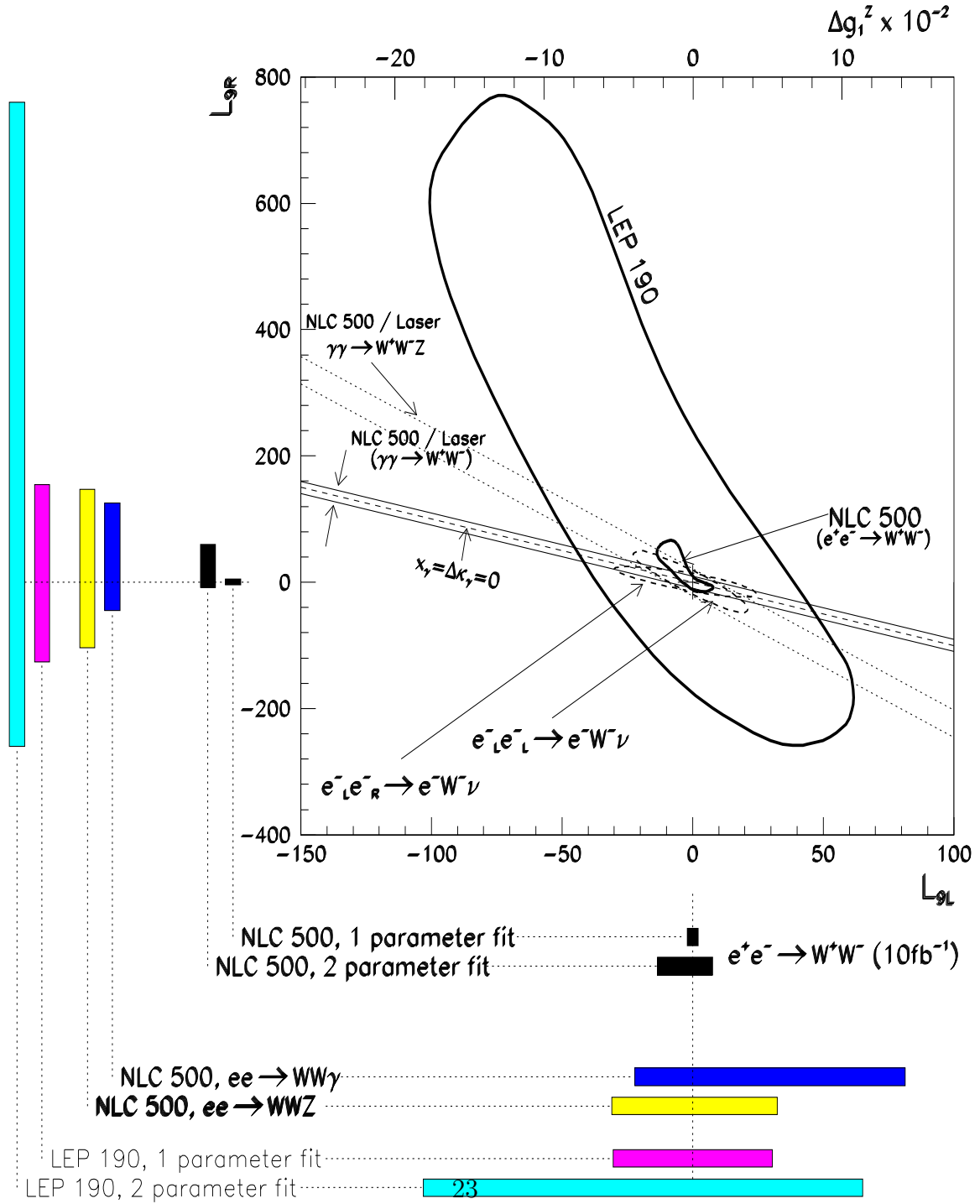}}
\end{center}
%[5 70 590 700]
\end{figure*}
The aim of the comparison is to show not only how well the different modes of the 
linear collider perform on bounding the parameters of the chiral Lagrangian 
but also how different fitting procedures constrain these couplings. In Fig.~\ref{l9fig} 
we show the results of $\chi^2$ fits, assuming in the case of a 500 GeV a total 
luminosity of only 10 fb$^{-1}$. The bounds from \epemt are from the 
BMT~\cite{BMT} analysis 
adapted to the chiral approach\footnote{We thank Misha~Bilenky for agreeing to 
conduct this analysis.}. The limit from \ememt are from an analysis by Frank Cuypers. 
The \gamgamt analysis is adapted from~\cite{Parisgg}. \\
\noi First one sees that there is 
an incredible improvement on the bounds expected from LEP2 when going
to 500 GeV. 
At this point a remark is in order. Complying with the benchmark values on the 
$L_i$ that one has set, the limits one expects to extract with a luminosity of 
$500$ pb$^{-1}$ at an energy of 190 GeV at LEP2 are not interesting from the point of 
view of symmetry breaking and precision. One only needs to contrast the situation with 
LEP1 and the equivalent plot of Fig.~\ref{Langacker.fig}. These limits, though better than 
those from the Tevatron, are still not meaningful even if one restricts to one parameter 
fits. Moving from two parameters to one, we gain a factor of three at
190 GeV. But still 
in this case one has $|L_{9L}|<30$ and $-125 <L_{9R}<155$. For such values the 
effective Lagrangian based on the chiral expansion is not meaningful and one should 
stick with the phenomenological parameterisation. As advertised the limits from 
$\gamma \gamma$ are quite competitive with those from \epemt at 500 GeV. The \ememt 
does not bring much information and does not usefully compete with the \epemt or 
even the \gamgamt options. Three vector production in \epemt, $WW\gamma$ and $WWZ$, 
are also not 
very helpful at 500 GeV from the point of view of checking the tri-linear couplings as 
indicated in the plot. 

\begin{figure*}[htb]
\caption{\label{newl9.fig}{\em Limits on ($L_{9L}-L_{9R}$) in 
\epemt including ISR and beam polarisation with only the resonant diagrams.
The effect of keeping all resonant diagrams for the semi-leptonic final state 
is also shown. Limits from $\gamma \gamma \ra W^+ W^-$ are also
included. The LHC limits from $WZ$ and $W^+W^-$ are inserted also. }}
%\cite{Barklow}
\begin{center}
%\mbox{\epsfxsize=8cm\epsfysize=8cm\epsffile{barklow.eps}}
%\epsfxsize=135mm
\mbox{\epsfysize=115mm\epsffile[18 163 531 645]{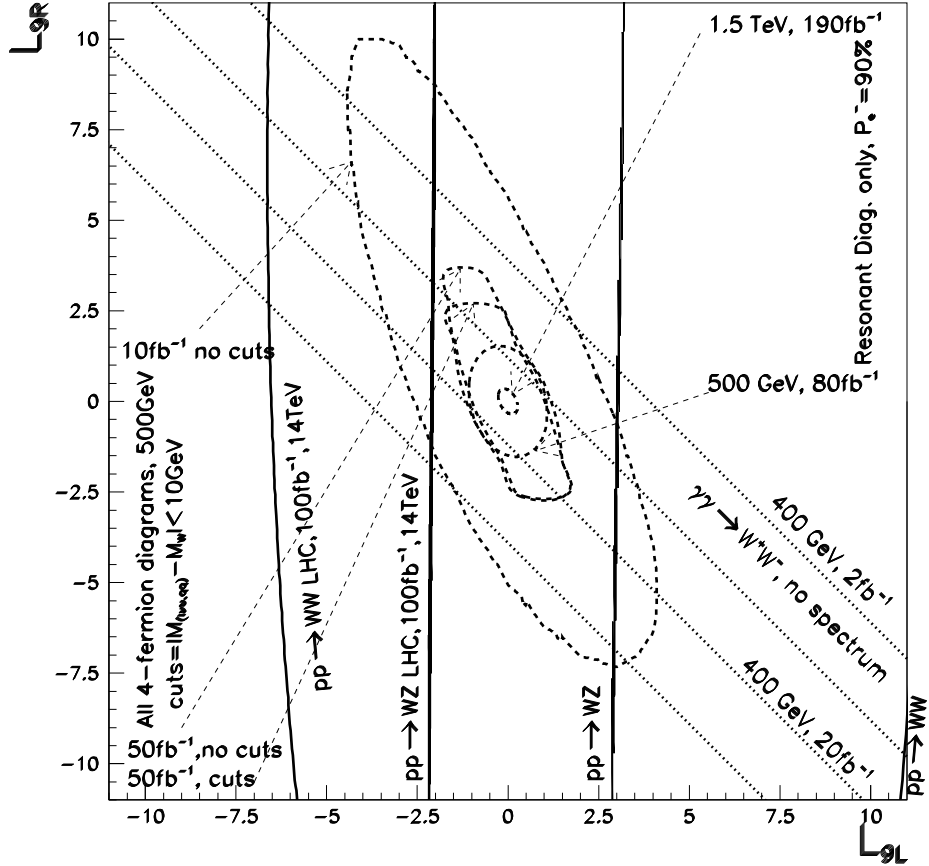}}
\end{center}
\end{figure*}

More recent analyses on the NLC have exploited the maximum likelihood technique and 
taken into account larger luminosities. First, the issue of the finite width can be 
quantified. Gintner, Godfrey and Couture~\cite{GGC} considered all diagrams that contribute to 
to the semi-leptonic $WW$ final state. By double mass constraint (10 GeV within $M_W$) 
one picks up essentially the double resonant process and compares this with the limits based 
on the same technique of fitting on all 5 kinematical variables. One sees, 
Fig.~\ref{newl9.fig}, that basing the analysis on the resonant diagrams very marginally 
degrades the limits. Therefore, after allowing for the $WW$ selection it seems that the limits
based on the resonant diagrams can be totally trusted. To a very good approximation changing the luminosity can be accounted 
for by a scaling factor, $\sim \sqrt{{\cal L}}$, compare with the same analysis 
done with a reduced luminosity of 10 fb$^{-1}$. The latter should also be compared with 
the analysis reported in the previous plot and which is based on a $\chi^2$ fit. As discussed earlier the 
maximum likelihood fit does better. 
This is confirmed also by the 
analysis conducted by Tim Barklow~\cite{Barklow} 
which assumes higher luminosities but take 
ISR into account as well as beam polarisation. For both the analyses
at 500 GeV and
1.5 TeV the luminosity shown on the plot is shared equally between a left-handed and 
a right-handed electron (assuming $90\%$ longitudinal electron polarisation). With 
80 fb$^{-1}$ (and almost similarly with 50 fb$^{-1}$) one really reaches the domain of 
precision measurements and one can truly contrast with the similar plot based on the 
LEP1 observables as concerns $L_{10}$ and $T_{new}$. It is quite fascinating that 
we can achieve this level of precision with such moderate energies. Moving to the TeV range 
one gains again as much as when moving from LEP190 GeV to 500 GeV, another 
order of magnitude improvement, see Fig.~\ref{newl9.fig}. 
Returning to 500 GeV it is clear that the $\gamma\gamma$ option really helps. 
The results shown in Fig.~\ref{newl9.fig} consider a luminosity of 20
fb$^{-1}$ with a peaked fixed spectrum corresponding to a cms \gamgamt
energy of 400 GeV. This set up helps reduce the contour of the 80
fb$^{-1}$-500 GeV analysis in the \epemt mode. 
This very promising result is arrived at solely on the basis of the
total cross section (with a moderate cut on the forward events). Only
statistical errors were taken into account. 
We expect a much better improvement when a full spin analysis is conducted and 
fits using the maximum likelihood method are performed. 
During this Workshop the Hiroshima group~\cite{Hiroshimagg} reported on a similar analysis 
with a full simulation. The 
electron energy was taken to be 250 GeV and the laser parameters were such 
that $x_0=4.7$. The original \epemt 
luminosity was assumed to be 
40 fb$^{-1}$ and a realistic \gamgamt simulated. 
These \gamgamt results also confirm the power of the \gamgamt mode and 
lead to the bound 
$|L_{9L}+L_{9R}| < 2.8$ at $90\%$CL. 

Fig.~\ref{newl9.fig} also compares the situation with the 
LHC. The limits on the parameters of the chiral Lagrangian have been adapted by Ulrich Baur 
from their detailed analysis that includes NLO QCD corrections, as 
outlined above when discussing the LHC. 
It is indeed found, as expected from the general arguments that
I exposed above, refer to Fig.~\ref{lkgfig}, that $L_{9L}$ is much better constrained 
than $L_{9R}$. As expected, the $WZ$ channel does a much better job since it does not 
suffer from the same ambiguity in the reconstruction of the final states and the 
background. Effectively $WZ$ (and hence $pp$) only constrains $L_{9L}$ through 
$\Delta g_1^Z$ that contributes a longitudinal $Z$. With 50 fb$^{-1}$
and 500 GeV 
the NLC constrains the two-parameter space much better than the LHC. The hadron 
collider is not very sensitive to $L_{9R}$. The sensitivity of the NLC is further enhanced 
if the experiments are done in conjunction with \ggwwt. 

What about the genuine quartic couplings, as parameterised through $L_{1,2}$. 
These are extremely important as they involve essentially solely the longitudinal modes
and hence are of crucial relevance when probing the Goldstone interaction. These are best 
probed through $V_L V_L \ra V_L V_L$ scattering. However, with only
500 GeV the 
$V_L$ luminosity inside an electron is unfortunately not enough and one has to 
revert to $\epem \ra W^+W^-Z$ production, as suggested in~\cite{nousee3v}. 
This has been taken up by 
A.~Miyamoto~\cite{Miyamoto} who conducted a detailed simulation including $b$ tagging to reduce 
the very large background from top pair production. With a luminosity of 
50 fb$^{-1}$ at 500 GeV, the limits are not very promising and do not pass the 
benchmark criterium $L_i<10$. It is found that $-95< L_1<71\;\;-103< L_2<100$ 
(one parameter fits). These limits agree very well with the results of a 
previous analysis~\cite{nousee3v}. 
To critically probe these special operators one needs energies in excess
of 1 TeV, a preliminary study at 1 TeV indicates that the bounds can improve to 
$L_{1,2} \sim 6$. However, it is difficult to beat the LHC here, where limits of order 
$1$ are possible~\cite{BDV} through $pp \ra W^+_L W^+_L $. 

In conclusion, it is clear that already with a 500 GeV \epemt collider
combined with a 
good integrated luminosity of about 50-80 fb$^{-1}$ one can reach a precision, 
on the parameters that probe \sb in the genuine tri-linear $WWV$ couplings, of
the same order as what we can be achieved with LEP1 on the two-point vertices. 
This would be an invaluable information on the mechanisms of symmetry breaking, 
if no particle has been observed at the LHC or ...the NLC (Light Higgs and 
SUSY). The NLC is particularly unique in probing the vector models 
$L_9$ with $L_{1,2}\sim 0 $ and hence is complementary to the LHC. The latter 
is extremely efficient 
at constraining the ``scalar" models. To probe deeper into the structure of 
symmetry breaking, a linear collider with an energy range 
$\sqrt{s}>$ 1.5 TeV would be 
most welcome. In this regime 
there is also the fascinating aspect of $W$ interactions that I have not
discussed 
and which is the appearance of strong resonances (refer too the talk of Tim 
Barklow). This would reveal another 
alternative to the \sm description of the scalar sector. 

{\bf Acknowledgements:}\\
 I am indebted to various friends and colleagues who have helped me 
with the talk by sending their results and answering many of my queries. 
I am particularly grateful to Ulrich~Baur, Kingman~Cheung, 
Frank~Cuypers, Markus~Gintner, 
Steve~Godfrey, George~Gounaris, Ken-ichi~Hikasa, Jean-Loic~Kneur, 
Yoshimasa~Kurihara, Klaus~Moenig, Akiya~Miyamoto, Robert~Sekulin, 
Rob~Szalapski 
and Tohru~Takahashi. I would also like to thank Marc 
Baillargeon and Genevi\`eve B\'elanger for the enjoyable collaboration. 
I am grateful to the organisers for the financial support of this 
excellently organised workshop. At last but not least, I would like 
to thank all the members of the Minami-Tateya group for a most enjoyable 
and memorable stay in Japan.

\end{document}